\documentclass[a4paper]{article}
\usepackage[utf8]{inputenc}
\usepackage[english]{babel} 
\usepackage{amsmath,amssymb} 
\usepackage{calrsfs}
\usepackage[normalem]{ulem}
\usepackage{comment}
\usepackage[colorlinks]{hyperref}
\usepackage[color=yellow]{todonotes}
\usepackage{arydshln}
\usepackage{graphicx}
\usepackage{esint}
\usepackage{booktabs}
\usepackage{chngpage}
\usepackage{graphicx}
\usepackage{enumerate}
\usepackage[margin=2cm]{geometry}
\usepackage{epstopdf}
\usepackage{listings}
\usepackage{lscape}
\usepackage{float, tikz, tikz-3dplot, tikz-cd, framed}
\usetikzlibrary{calc}
\usepackage[colorlinks]{hyperref}
\usepackage{pdfpages}
\usepackage{fancyhdr} 
\hypersetup{colorlinks={true},linkcolor={black}}
\DeclareMathAlphabet{\pazocal}{OMS}{zplm}{m}{n}
\newtheorem{theorem}{Theorem}

\newtheorem{definition}[theorem]{Definition}

\newtheorem{lemma}[theorem]{Lemma}

\newtheorem{remark}[theorem]{Remark}

\numberwithin{theorem}{section}

\newcommand{\scp}[2]{{\big\langle {#1}\, , \, {#2}\big\rangle}}
\newcommand{\Scp}[2]{{\Big\langle {#1}\, , \, {#2}\Big\rangle}}
\newcommand{\SCP}[2]{{\left\langle {#1}\, , \, {#2}\right\rangle}}

\newcommand{\ob}[1]{\overline{#1}}
\newcommand{\wt}[1]{\widetilde{#1}}
\newcommand{\bm}[1]{\mbox{\boldmath{$#1$}}}

\def\a{{\alpha}}
\def\e{{\epsilon}}
\def\eps{\varepsilon}
\def\om{{\omega}}

\def\bxi{{\bm \xi}}
\def\q{{\bf{q}}}
\def\p{{\partial}}
\def\l{{\bf{x}_0}}
\def\bx{{\bf x}}
\def\bX{{\bf X}}
\def\ba{{\bf a}}
\def\bk{{\bf k}}
\def\bfm{{\bf m}}
\def\bR{{\bf R}}
\def\bu{{\bf u}}
\def\bU{{\bf U}}
\def\bv{{\bf v}}
\def\u{{\bf{u}}}

\def\de{{\delta}}
\def\L{{\cal L}}

\def\Hbar{\ob{H}}
\def\bb{\ob{b}}
\def\mb{\ob{\bf m}}
\def\pb{\ob{\bf p}}
\def\pbL{\ob{p}^L}
\def\ub{\ob{\bf u}}
\def\Rb{\ob{\bf R}}

\def\PhibL{{\ob{\Phi}^L}}

\def\bxi{{\bm{\xi}}}
\def\bzeta{{\bm{\zeta}}}

\def\omt{\tilde\omega}
\def\Dt{\widetilde{D}}

\def\contract{\makebox[1.2em][c]{\mbox{\rule{.6em}
{.01truein}\rule{.01truein}{.6em}}}}

\DeclareMathOperator{\diff}{d\!}

\numberwithin{equation}{section} 
\numberwithin{figure}{section} 
\numberwithin{table}{section} 

\makeatletter
\renewenvironment{framed}{%
 \def\FrameCommand##1{\hskip\@totalleftmargin
 \fboxsep=\FrameSep\fbox{##1}
     \hskip-\linewidth \hskip-\@totalleftmargin \hskip\columnwidth}%
 \MakeFramed {\advance\hsize-\width
   \@totalleftmargin\z@ \linewidth\hsize
   \@setminipage}}%
 {\par\unskip\endMakeFramed}
\makeatother

\begin{document}
\title{Stochastic Closures for Wave--Current Interaction Dynamics
}
\author{Darryl D. Holm 
\\Mathematics Department 
\\Imperial College London
\\ \normalsize email: d.holm@ic.ac.uk
}

\date{} 

\maketitle

\makeatother

\begin{abstract}

 Wave--current interaction (WCI) dynamics energizes and mixes the ocean thermocline by producing a combination of Langmuir circulation, internal waves and turbulent shear flows, which interact over a wide range of time scales. Two complementary  approaches exist for approximating different aspects of WCI dynamics.  These are the Generalized Lagrangian Mean (GLM) approach and the Gent--McWilliams (GM) approach. 
Their complementarity is evident in their Kelvin circulation theorems. GLM introduces a wave pseudomomentum per unit mass into its Kelvin circulation integrand, while GM introduces a an additional `bolus velocity' to transport its Kelvin circulation loop. The GLM approach models Eulerian momentum, while the GM approach models Lagrangian transport. 
In principle, both GLM and GM are based on the Euler--Boussinesq (EB) equations for an incompressible, stratified, rotating flow. The differences in their Kelvin theorems arise from differences in how they model the flow map in the Lagrangian for the Hamilton variational principle underlying the EB equations. 

A recently developed approach for uncertainty quantification in fluid dynamics constrains fluid variational principles to require that Lagrangian trajectories undergo Stochastic Advection by Lie Transport (SALT). Here we introduce stochastic closure strategies for quantifying uncertainty in WCI by adapting the SALT approach to both the GLM and GM approximations of the EB variational principle. In the GLM framework, we introduce a stochastic group velocity for transport of wave properties, relative to the frame of motion of the Lagrangian mean flow velocity and a stochastic pressure contribution from the fluctuating kinetic energy. In the GM framework we introduce a stochastic bolus velocity in addition to the mean drift velocity by imposing the SALT constraint in the GM variational principle.

\end{abstract}


\tableofcontents

\section{Introduction}\label{intro-sec}

The wind drives gravity waves on the ocean surface. Over time, the collective action of these wind-driven gravity waves on the ocean surface generates Langmuir circulations (LC) which transport heat and mix material properties deeper into the ocean. The presence of LC is seen as ``lines on the sea surface" marked by flotsam trapped between roughly parallel, horizontally counter-circulating pairs of Langmuir vortex rolls. Eventually, these wave-current interactions energise and mix the ocean surface boundary layers (OSBLs) which occupy the upper few hundred meters of the ocean. In turn, the well-mixed region at the top of the OSBLs comprises the thermocline. Just below it the stratified regions propagate internal waves which further transmit and disperse wave activity. 

The turbulent wave-current mixing by Langmuir circulation seen in the OSBL is important in climate modelling, because it controls the exchange of heat and trace gases between the atmosphere and ocean through the mix layer. However, a difficulty arises in numerically simulating the regional effects of Langmuir circulation on turbulent mixing in OSBL, because of the huge disparities among length and times scales of the waves, currents, regional flows and their effects on climate.  Such huge disparities make direct numerical simulations (DNS) of turbulent mixing by wave, current interaction intractable for any existing or projected computer for decades to come. 

For comprehensive reviews of modern approaches for quantifying the dynamics of Lagrangian flows such as Langmuir circulations coupled to surface and internal waves, see, e.g., Sullivan and McWilliams \cite{SuMcW2010}, Phillips [2003] \cite{Phillips-LC2003}, Fujiwara et al. [2018] \cite{Fujiwara-etal2018} as well as references therein. 

Current parameterizations of turbulent mixing in numerical simulations of the OSBL lead to substantial systematic errors, for example, in predicting the depth of the OSBL for a given wind stress. These errors, in turn, lead to further uncertainty in predictions of sea surface temperature and rate of exchange of gases such as $CO_2$ between the ocean and the atmosphere, \cite{Belcher2012}. 

Because of the computational intractability due to the enormous scale disparity and the space-time distributed nature of wave-current interactions with weather and climate dynamics, simulations of turbulent mixing in OSBL are always carried out in regions of parameter space which are far from the observed values, either with: (a) an unphysical lack of scale separation between the energy-containing, inertial, and dissipative scales while parameterizing the missing physics, or with (b) a study of the processes at much smaller length scales, often with periodic boundaries (unphysical at large scales but used under the hypothesis of spatial homogeneity of the flows). 
Moreover, because of the nonlinear nature of turbulent flows and the ensuing multi-scale, space-time distributed interactions, the physics of the unresolvable, rapid, small scales may differ significantly from the properties (e.g., statistics) of the resolvable large scales. For example, the regime of asymptotic expansions for the large scale computational models occurs at small Rossby number, which enforces hydrostatic and geostrophic balances. However, for wave--current interaction (WCI) at the submesoscale length scales below the Rossby radius where Langmuir circulations develop, the Rossby number is order $O(1)$ and neither of these large-scale balances is enforced. This imbalance requires another model.

Given this situation, there is clearly a need for enhanced methods for parameterizing the effects on the resolvable scales of the unresolvable small scales in space and time. Two main parameterization approaches have been developed over the years to model the effects of the unresolvable small scales in turbulence on the scales resolved in the simulations. 

The first parameterization approach is primarily computational, via Large Eddy Simulations (LES). LES is widely used in engineering, in atmospheric sciences, and to a lesser extent in astrophysics. However, in the LES approach, many important physical parameters for the Langmuir circulations are not scale--appropriate. For example, in the LES approach, the Reynolds number $Re$ is not known at the Langmuir scale. Instead, one may attempt modelling the behavior of the Langmuir flow in the limit that $Re$ is very large. LES is an important tool for phenomenological discovery and quantification in wave-current interactions. However, it is known to be vulnerable to significant uncertainty in its sub-grid-scale modelling  \cite{SuMcW2010,Pearson-etal-2017}. For a comprehensive review of parameterization in computational ocean modelling, see \cite{HCDF-K-ocean2017}. For considerations of LES design for computational studies of  global ocean circulation, see \cite{HHPW}.

The second parameterization approach is primarily theoretical. The theory is traditionally based on the work of Craik \cite{CL,C82a,C82b,C85} with later extensions by Leibovich \cite{L2,L3,L4,LT}. In the Craik-Leibovich (CL) model of Langmuir circulation, wave-induced fluid motion affects the OSBL at local scales via the `Stokes mean drift velocity'  through a `vortex force' as well as material advection. These two effects combine to produce the instability which creates the Langmuir circulation. 

In WCI, the waves are propagating through the moving  fluid at a speed comparable to the fluid velocity itself. This means that the wave frequency is Doppler shifted by the fluid motion. However, the wave interaction is by no means frozen into the fluid motion.  Instead, the wave--current interaction (WCI) is distributed along the path of the wave through the comparably moving  fluid. In particular, the Eulerian mean group velocity of the wave is defined relative to the frame moving with fluid \cite{HolmCL1996}, and the Eulerian-mean WCI dynamics at a given fixed point in space depends on the history of wave interaction all along the entire Lagrangian path of the fluid parcel currently occupying that point. Mathematically, this implies that the description of WCI must be formulated in terms of the Eulerian mean of the \emph{pull-back} of the fluid properties by the Lagrange-to-Euler map, which is assumed to be a smooth invertible map.  This is a hybrid description in which the wave activity takes place in the frame of motion of the fluid. 

The WCI situation is addressed directly by the Generalized Lagrangian Mean (GLM) approach formulated in Andrews \& McIntyre \cite{AM1978a,AM1978b}. GLM generalizes the CL approach by decomposing the flow into its fast and slow components, then taking various types of time-averages, phase-averages and asymptotic approximations of the wave--current interaction dynamics at which the Rossby number is order $O(1)$. In GLM, another dynamical variable is introduced, called the pseudomomentum, in addition to the Stokes mean drift velocity appearing in the CL approach. Thus, in GLM, the fast-slow split in time is performed at a single spatial scale. No differences in spatial scale of the waves and mean flow need to be assumed.  Relevant references relevant to our purposes here are \cite{AM1978a,GjHo1996,GiVa2018}. 


\paragraph{Aims of the paper.}
This paper aims to lay down a mathematical foundation which has the potential for both quantifying and reducing the uncertainties in the numerical simulation of ocean-atmosphere mixing layer dynamics, by developing new methods of enhanced modelling of sub-grid-scale (SGS) circulation effects in the OSBL produced by wave--current interactions (WCI). Our approach is based on structure-preserving approaches in data-driven stochastic modelling for quantifying these uncertainties, combined with data assimilation methods for reducing uncertainty.  Recent applications of this  approach for data analysis and simulation for two-dimensional confined fluid flows are reported in \cite{CoCrHoPaSh2018a,CoCrHoPaSh2018b}. Specifically, we lay foundations for extending the approach of \cite{Holm2015, HolmRT2018, CrFlHo2018, CoCrHoPaSh2018a, CoCrHoPaSh2018b} from incompressible flows in fixed domains to incompressible rotating stratified flows driven by sub-grid-scale dynamics represented by stochastic processes in three dimensions. Our approach via averaged variational principles is designed to preserve the fundamental nonlinear structure of fluid dynamics. Above all, it introduces stochasticity while preserving the nature of fluid transport, the Kelvin circulation theorem and the geometric structure of fluid dynamics, including its Lie--Poisson Hamiltonian structure. In particular, our approach takes advantage of the recent developments in stochastic fluid dynamics based on geometric mechanics in \cite{Holm2015,HolmRT2018,BdLHoLuTa2019,DrHo2019} to introduce a stochastic closure procedure which preserves the geometrical structure of GLM.

The present paper also provides the derivation of a certain stochastic wave--current interaction (SWCI) model. The SWCI model is based on a stochastic closure of the well-known GLM description of the Euler--Boussinesq (EB) equations for a rotating, stratified, incompressible fluid flow. Its derivation is based on GLM averaging of a constrained Hamilton's principle for the EB equations in the Eulerian representation, leading to Euler--Poincar\'e variational equations for the GLM description, coupled to an Eulerian mean description of the fluctuation dynamics. This formulation is developed via a Legendre transformation into a Lie--Poisson Hamiltonian description, \cite{HoKu1983,HMR1998}. 

In this Hamiltonian setting, two natural stochastic closures of the GLM theory present themselves. The first closure assumes that the unknown GLM group velocity and the GLM kinematic pressure in the Hamiltonian are each temporally stochastic in the Stratonovich sense, with separate stationary spatial correlations. This closure amounts to a stochastic parameterization of the GLM group velocity and the GLM kinematic pressure whose spatial correlations must be calibrated from observed or simulated data. However, this data for the GLM stochastic closure appears to be rather inaccessible.

The elusiveness of data for the two GLM wave closures suggests the formulation of an alternative closure which directly separates the time scales of the fluid transport velocity into its slow fluid and fast wave parts. This approach is reminiscent of the introduction of the \emph{bolus velocity} in the celebrated Gent-McWilliams (GM) parameterization of subgrid-scale transport  \cite{Gent2011,GM1990,GM1996}, which is generally used in computational simulations of ocean circulation. 

After discovering the elusiveness of the data required in formulating the stochastic WCI closure for GLM, in the second part of the paper, we propose an alternative stochastic closure for WCI.  The alternative stochastic closure proposed here is a variant  of the existing theory in \cite{FGB-Ho2018} of Stochastic Advection by Lie Transport (SALT) \cite{Holm2015,HolmRT2018,CoGoHo2017,CrFlHo2018} which introduces Hamiltonian stochastic transport into the material fluid evolution as a constraint in Hamilton's variational principle for fluid dynamics. The SALT approach separates the slow and fast time scales of the fluid transport velocity into drift and stochastic parts. Implementation of this closure has already been tested in \cite{CoCrHoPaSh2018a, CoCrHoPaSh2018b} and found to be quite accessible for calibration by observational data from both high resolution computational simulations. Because it deals with enhanced transport, the SALT approach is naturally compatible with formulating a data-driven stochastic version of GM parameterization of transport by unresolved time scales. 

Stochastic parameterizations have been commonly used in both atmosphere and ocean sciences, ever since the break-through results of \cite{Buizza-etal1999}. Indeed, other stochastic versions of the GM already exist, as reviewed in \cite{GrKl2019}, and the future comparisons of these approaches with the two stochastic approaches presented here for GLM and GM are bound to be interesting. 

\paragraph{Plan.}
In section \ref{GLM-rev-sec} we will review some background information from the GLM theory relevant to the remainder of the paper. We refer to Appendix \ref{EBfluid-sec} for the details in deriving the deterministic GLM equations for the Euler--Boussinesq equations in the Euler-Poincar\'e variational framework \cite{HMR1998}, and the passage to the Lie--Poisson Hamiltonian side as an arena for seeking a natural stochastic closure.

Section \ref{SPDEs-sec} introduces stochastic closures for the GLM equations. 

By way of prepparation, Section \ref{StochTrans-sec} provides a summary of the Kunita--It\^o--Wentzell (KIW) theorem, which proves the key formula in stochastic transport. Section \ref{SPDEs-sec} then uses the KIW formula to  investigate stochastic closures of the GLM Euler--Boussinesq equations due to both pressure and displacement fluctuations. Section \ref{SPDEs-sec} also advocates an alternative closure based on taking Stochastic Advection by Lie Transport (SALT) as a general strategy, rather than proliferating the possible sources of stochasticity for the various types of subgrid-scale physics for which our knowledge is incomplete. 

In Section \ref{GMapproach-sec}, the Gent--McWilliams (GM) transport scheme is reviewed and adapted to the variational SALT strategy in \cite{FGB-Ho2018}. 

Section \ref{conclude-sec} summarizes our conclusions and outlook for open problems.


\section{Brief review of GLM theory for Euler--Boussinesq fluids}
\label{GLM-rev-sec}

The Generalized Lagrangian Mean (GLM) theory of
nonlinear waves on a Lagrangian-mean flow is formulated in two
consecutive papers of Andrews \& McIntyre [1978a,b] \cite{AM1978a,AM1978b}. The present
section reviews what we shall need later from the rather complete
description given in these papers. See also the textbook by B\"uhler \cite{Buhler2014} for an accessible
update on the GLM theory. Even now, these fundamental papers still make worthwhile reading and 
they are taught in many atmospheric science departments, because 
they represent an exceptional accomplishment in formulating 
averaged motion equations for fluid dynamics.


\subsection{Relevant information from the GLM theory}\label{GLM-bckgrnd-sec}


\subsubsection{Defining relations for Lagrangian mean
\& Stokes correction in terms of Eulerian mean}

The GLM equations are based on defining fluid quantities at a
displaced fluctuating position $\mathbf{x}^\xi := \mathbf{x}+\xi(\mathbf{x},t)$. 
In the GLM description, $\overline{\chi}$
denotes the Eulerian mean of a fluid quantity
$\chi=\overline{\chi}+\chi^{\,\prime}$ while $\overline{\chi}^L$ denotes the
Lagrangian mean of the same quantity, defined by
\begin{equation}
\overline{\chi}^L(\mathbf{x})
\equiv
\overline{\chi^\xi(\mathbf{x})}
\,,\quad\hbox{with}\quad
\chi^\xi(\mathbf{x})
\equiv
\chi(\mathbf{x}+\xi(\mathbf{x},t))
\,.
\label{LM-def-rel}
\end{equation}
Here $\mathbf{x}^\xi\equiv\mathbf{x}+\xi(\mathbf{x},t)$ is the current
position of a Lagrangian fluid trajectory whose current mean position is 
$\mathbf{x}$. Thus, $\xi(\mathbf{x},t)$ with vanishing Eulerian 
mean $\overline{\xi}=0$ denotes the fluctuating displacement of a Lagrangian
particle trajectory about its current mean position $\mathbf{x}$. 

\begin{remark}\rm
Fortunately, this GLM notation is also {\it standard} in the
stability analysis of fluid equilibria in the Lagrangian picture.
See, e.g., the classic works of Bernstein et al. [1958], Frieman \&
Rotenberg [1960] and Newcomb [1962]. See Jeffrey \& Taniuti [1966]
for a collection of reprints showing applications of this approach
in controlled thermonuclear fusion research. For insightful reviews,
see Bernstein [1983], Chandrasekhar [1987] and, more recently, Hameiri
[1998]. Rather than causing confusion, this confluence of notation
encourages the transfer of ideas between traditional Lagrangian
stability analysis for fluids and GLM theory.
\end{remark}

In GLM theory, the difference  $\chi^\xi-\overline{\chi}^L=\chi^\ell$ is
called the {\bf Lagrangian disturbance} of the quantity $\chi$. One
finds $\overline{\chi^\ell}=0$, since the Eulerian mean possesses the
{\bf projection property} $\overline{\overline{\chi}}=\overline{\chi}$ for any
quantity $\chi$ (and, in particular, it possesses that property
for $\chi^\xi$).%
\footnote{Note that spatial filtering in general does {\it not}
possess the projection property.}
Andrews \& McIntyre [1978a] \cite{AM1978a} show that, provided the
smooth map $\mathbf{x}\to\mathbf{x}+\xi(\mathbf{x},t)$ is invertible
(that is, provided the vector field $\xi(\mathbf{x},t)$ generates a
diffeomorphism), then the Lagrangian disturbance velocity
$\mathbf{u}^\ell$ may be expressed in terms of $\xi$ by
\begin{equation}\label{u-ell-def}
\mathbf{u}^\ell
=
\mathbf{u}^\xi
-
\overline{\mathbf{u}}^L
=
\frac{D^L\xi}{Dt}
\,,\quad\hbox{where}\quad
\frac{D^L\xi}{Dt}
\equiv
\frac{\partial\xi}{\partial t}
+
\overline{\mathbf{u}}^L\cdot\nabla\xi
\,.
\end{equation}
Consequently, the Lagrangian disturbance velocity $\mathbf{u}^\ell$ is
a genuine fluctuation quantity satisfying
$\overline{\mathbf{u}^\ell}=0$, since
$\overline{\mathbf{u}^\xi} - \overline{\overline{\mathbf{u}}^L} 
= \overline{\mathbf{u}^\xi} - \overline{\overline{\mathbf{u}^\xi}}
=0$, by the projection property. Alternatively,
$\overline{\mathbf{u}^\ell}=\overline{D^L\xi/Dt}=0$ also follows,
since the Eulerian mean commutes with ${D^L}/{Dt}$ and $\xi(\mathbf{x},t)$ has
mean zero.

To summarise, GLM sets $ \mathbf{u}^\xi(\mathbf{x},t) := \mathbf{u} (\mathbf{x} + {\bxi}(\mathbf{x},t))$ 
where $\mathbf{x}$ is evaluated as the current position on a Lagrangian mean path and 
\begin{equation} 
\mathbf{u}^\xi := \frac{D^L}{Dt}\Big(\bx+\bxi(\bx,t)\Big) 
= \overline{\mathbf{u}}^L(\bx,t) + \mathbf{u}^\ell(\bx,t)
\quad\hbox{with}\quad
\frac{D^L}{Dt} = \frac{\p}{\p t}  + \mathbf{u}^L\cdot \frac{\p}{\p \bx}
\quad\hbox{and}\quad
\mathbf{u}^\ell := \frac{D^L{\xi}}{Dt}
\,.
\label{u(l)}
\end{equation} 
One then defines the Lagrangian mean velocity as $\ob{\mathbf{u}^\xi}(\bx,t) = \overline{\mathbf{u}}^L(\bx,t)$, 
where $\overline{(\,\cdot\,)}$ is a time, or phase average at fixed Eulerian coordinate $\mathbf{x}$. 



\subsubsection{The pull-back representation of fluctuations in fluid motion}

Here we briefly explain the GLM approach from the viewpoint of \cite{CoGoHo2017}, whose  multi-time homogenization analysis led to a stochastic formulation of the type proposed in the present paper. We will use the slightly expanded notation of \cite{CoGoHo2017} in this remark and then revert later to GLM notation. For this purpose, we will need to employ the action on functions $f$, $k$-forms $\alpha$ and vector fields $X$ of smooth maps $\phi$ via \emph{pull-back}, denoted $\phi^*$ and defined as the composition of functional dependence from the right. For example, the expression 
\[
\phi^*f := f \circ \phi
\,,\]
is called the pull-back of the function $f$ by the smooth map $\phi$. This notation will also be applied to $k$-forms and vector fields. The inverse of the pull-back is called the \emph{push-forward}. It is the pull-back by the inverse map.

The GLM theory can be described \cite{CoGoHo2017} as the Eulerian mean with respect to fast time dependence of the \textit{pull-back} of the fluid properties by an evolutionary fluid flow map $g_t = \widetilde{g}_{t/\eps}\circ \ob g_t $ with two time scales, one slow and one fast. This map is defined as the composition of a mean flow map $\ob g_t$ depending on slow time $t$ and a rapidly fluctuating flow map $\widetilde{g}_{t/\eps}$ associated with the evolution of the fast time scales $t/\eps$, with $\eps\ll1$. The GLM notation is recovered by defining the flow map associated with the fast scales as the (spatially) smooth invertible map with smooth inverse (i.e., a diffeomorphism, or diffeo for short) given by the \emph{sum},
\begin{align}
\widetilde{g}_{t/\eps} = \operatorname{Id} + \,{\zeta}_{t/\eps}
\quad\hbox{where}\quad \eps\ll1
\, .
\label{e.gprime}
\end{align}
The full flow map is taken to be the composition of $\ob g_t$ and $\widetilde{g}_{t/\eps}$, as
\begin{align}
g_t = \widetilde{g}_{t/\eps}\circ \ob g_t = \ob g_t + {\zeta}_{t/\eps}\circ \ob g_t\, .
\label{compmap}
\end{align}
The Lagrangian trajectory of a fluid parcel is then given by $\q(\l,t)=g_t\l$, so that
\begin{align}
\q(\l,t)=g_t\l
\quad\Longrightarrow\quad
\q(\l,t) = \ob{\q}(\l,t) + {\zeta}_{t/\eps}\circ \ob{\q}(\l,t)\,,
\label{e.q}
\end{align}
where the vector $\l$ denotes the fluid label, e.g., the initial condition of a fluid parcel. 

Equation \eqref{e.q} is equivalent to the displaced fluctuating position denoted as $\mathbf{x}^\xi := \mathbf{x}+\xi(\mathbf{x},t)$, in the GLM notation. That is, the rapidly fluctuating vector displacement field 
\begin{align}
\bm{\zeta}(\ob{\q}(\l,t),{t/\eps}):={\zeta}_{t/\eps}\circ \ob{\q}(\l,t)
\label{e.vdf}
\end{align}
is defined along the slow, large-scale, resolved trajectory, $\ob\q$. At this point, (\ref{e.q}) may be taken as exact, since it follows directly from the definition of the map ${\zeta}_{t/\eps}$ in (\ref{e.gprime}). Thus, we have
\begin{align}
\q(\l,t) = \ob{\q}(\l,t) + \bm{\zeta}(\ob{\q}(\l,t),{t/\eps})\,.
\label{zeta-q0}
\end{align}

The tangent to the composite flow map $g_t$ in \eqref{compmap} at $\q(\l,t)$ along the Lagrangian trajectory \eqref{e.q} defines the Eulerian velocity vector field $\u$, written as
\begin{align}
\dot{g}_t\l = {\dot{\q}}(\l,t) = \u(\q(\l,t),t)\, .
\end{align}
Differentiation of the Lagrangian trajectory \eqref{zeta-q0} including the assumed fluctuating displacement field (\ref{e.vdf}) yields 
\begin{align}
\u(\q(\l,t),t)
&=
\u(\ob\q +\zeta_{t/\eps}\circ \ob \q,t) 
\\
= {\dot{\q}}(\l,t)
&= {\dot{\ob{\q}}}
+  ({\dot{\ob{\q}}}\cdot {\bm{\nabla}_{\ob\q}})\,\bzeta(\ob\q(\l,t),t/\eps)
+ \frac{1}{\eps}\, \partial_{t/\eps}\bzeta \, .
\label{e.qdot}
\end{align}
This is equivalent to the definition of $\mathbf{u}^\xi$ in equation \eqref{u(l)}, in the GLM notation. See  \cite{CoGoHo2017} for more discussion of the pull-back representation of fluctuations in fluid dynamics, including
results of multi-time homogenisation leading to a stochastic representation of the Lagrangian trajectory in the limit that the ratio of slow and fast time scales diverges. In this case, the decomposition \eqref{compmap} becomes a composition of a stochastic map and a deterministic map. 

\subsubsection{Summary of natural operations on differential $k$-forms ($\Lambda^k$)}
Differential forms are objects you can integrate. Manifolds are spaces on which the rules of calculus apply. A $k$-form $\alpha\in\Lambda^k$ on a smooth manifold $M$ is defined by the antisymmetric wedge product of $k$ differential basis elements, as
\[
\alpha = \alpha _{{i_1}\dots{i_k}}(x) dx^{i_1}\wedge \dots \wedge dx^{i_k} \in\Lambda^k(M)\,,
\]
in which the function $\alpha _{{i_1}\dots{i_k}}(x)$ is totally antisymmetric under exchange of any two neighbouring indices. 
Three basic operations are commonly applied to differential forms defined on a smooth manifold, M. The three operations are: exterior derivative (${\rm d}$), contraction ($\contract$) and Lie derivative ($\pounds_X$) in the direction of a vector field $X$. These three operations act as follows. 

$\bullet$\quad\rm Exterior derivative ($d\alpha$) raises the degree:
$
d\Lambda^k \mapsto  \Lambda^{k+1}
\,.
$

$\bullet$\quad\rm Contraction ($X \contract\alpha$) with a vector field $X$ lowers the degree:
$
X\contract \Lambda^k \mapsto  \Lambda^{k-1}
\,.
$

$\bullet$\quad\rm Lie derivative ($\pounds_X\alpha$) by vector field $X$ preserves the degree,
$
\pounds_X\Lambda^k \mapsto  \Lambda^k.
$

\begin{remark}\rm$\,$\\
For a $k$-form $\alpha\in\Lambda^k$, the Lie derivative $\pounds_X\alpha$ is defined geometrically by
\begin{eqnarray*}
\pounds_X\alpha= X \contract {\rm d}\alpha + {\rm d} ( X \contract \alpha)
\,.
\end{eqnarray*}
This geometric definition of the Lie derivative is called {\bf Cartan's formula}. 

Note that the Lie derivative commutes with the exterior derivative. That is,
\begin{eqnarray*}
{\rm d}(\pounds_X\alpha) = \pounds_X {\rm d}\alpha
\,,
\quad\hbox{for}\quad
\alpha\in\Lambda^k(M)
\quad\hbox{and}\quad
X\in\mathfrak{X}(M)
\,.
\end{eqnarray*}
This useful property may be proved via a direct calculation which uses Cartan's formula and the property of the exterior derivative ${\rm d}$ that ${\rm d}^2=0$.
\end{remark}

\subsubsection{How pull-back dynamics leads to Lie derivatives}

The pull-back $\phi_t^*$ of a spatially smooth flow $\phi_t$ on a smooth manifold $M$ generated by a smooth vector field $X\in\mathfrak{X}(M)$ commutes with the exterior derivative ${\rm d}$, wedge product $\wedge$ and contraction $\contract$. 
For an introduction to geometric fluid mechanics based on these standard concepts, see \cite{Holm2011}.

A smooth time-dependent invertible map with a smooth inverse (i.e., a diffeomorphism) $\phi_t\in Diff(M)$ acting on a smooth manifold $M$ may be generated by integration along the characteristic curves of a smooth vector field $X(\bx,t)\in \mathfrak{X}(M)$ via $d{\phi}_t/dt = X_t \circ \phi_t$. Under the action of such a smooth invertible map $\phi_t$ on $k$-forms $\alpha,\,\beta\in\Lambda^k(M)$, at a point $\bx\in M$, the pull-back $\phi_t^*$ is natural for the three operations ${\rm d}$, $\wedge$ and $\contract$. That is,
\begin{align*}
{\rm d}(\phi_t^*\alpha)
&=
\phi_t^*{\rm d}\alpha
\,,\\
\phi_t^*(\alpha\wedge\beta)
&=
\phi_t^*\alpha\wedge\phi_t^*\beta
\,,
\\
\phi_t^*(X\contract \alpha)
&=
\phi_t^*X\contract \phi_t^*\alpha
\,.
\end{align*}
In addition, the Lie derivative $\pounds_X\alpha$ of a $k$-form $\alpha\in\Lambda^k(M)$ by the vector field $X$ tangent to the flow $\phi_t$ on $M$ with $\phi_t|_{t=0}=Id$ may be defined either dynamically or geometrically (by Cartan's formula) as
\begin{align}
\pounds_X\alpha
&=
 \frac{d}{dt}\bigg|_{t=0}(\phi_t^*\alpha)
=
X \contract {\rm d}\alpha
+ {\rm d}( X \contract \alpha)
,
\label{Dyn+Geom-def-LieDeriv}
\end{align}
in which the last equality in (\ref{Dyn+Geom-def-LieDeriv}) is Cartan's geometric formula  for the Lie derivative. The  equivalence of the dynamic and geometric definitions of the Lie derivative in the last equality may be proved directly. This equivalence can be quite informative. For example, in the case $\alpha(\bx) = u_i(\bx){\rm d}x^i$ for $\bx\in \mathbb{R}^3$, $i=1,2,3,$ this equivalence implies a well-known vector calculus identity, namely
\begin{align}
\begin{split}
\pounds_X(u_i(x){\rm d}x^i)
:=
 \frac{d}{dt}\bigg|_{t=0}\phi_t^*(u_i(x){\rm d}x^i)
 &=
\bigg[\frac{\partial u_i (\phi_t(x))}{\partial \phi_t^j(x)}  \frac{d\phi_t^j}{dt}\bigg]_{t=0}\hspace{-2mm}dx^i
+ u_i(x) {\rm d} \bigg[\frac{d}{dt} \phi_t^j(x)\bigg]_{t=0}
\\&= \bigg[ \frac{\partial u_i (x)}{\partial x^j}X^j   + u_j(x) \frac{\partial X^j (x)}{\partial x^i} \bigg] {\rm d}x^i
\\&= \big[ (\bX\cdot\nabla) \bu + u_j \nabla X^j \big]\cdot {\rm d}\bx
\,,\\
X \contract {\rm d}(\bu \cdot {\rm d}\bx)
+ {\rm d} \big( X \contract (\bu \cdot {\rm d}\bx)\big)
&=
 \big[ - \bX \times {\rm curl}\bu + \nabla ( \bX\cdot \bu) \big]\cdot {\rm d}\bx
 \,.
 \end{split}
\label{FIFD}
\end{align}
The equality of these two expression, of course, yields the fundamental vector calculus identity of fluid dynamics. This calculation turns out to be the basis of the Kelvin circulation theorem.
\begin{definition}\hskip-2pt
{\bf(Pull-back and push-forward Lie derivative formulas)}\rm$\,$\\
The mathematical basis for analysis of fluid transport is the following text-book formula \cite{MaHu1983} which relates the pull-back to the Lie derivative: 
\begin{align}
 \frac{d}{dt}(\phi_t^*\alpha)
 =
 \phi_t^*\big(\p_t\alpha + \pounds_X\alpha\big)
 \,.
\label{Pull-back-comp}
\end{align}
In words, the tangent to the pull-back $\phi_t^*\alpha$ of a time dependent differential $k$-form $\alpha\in\Lambda^k(M)$ by a smooth invertible flow map $\phi_t$ is the pull-back $\phi_t^*$ of the Lie derivative of  the $k$-form $\alpha$ with respect to the vector field $X$ that generates the flow, $\phi_t$. 

Likewise, for the push-forward, which is the pull-back by the inverse, $(\phi_t)_*=(\phi_t^{-1})^*$, we have
\begin{align*}
 \frac{d}{dt}((\phi_t^{-1})^*\alpha_0)
 =
-(\phi_t^{-1})^*\big(\pounds_X\alpha_0\big)
 \,,
\end{align*}
or, equivalently,
\begin{align}
 \frac{d}{dt}((\phi_t)_*\alpha_0)
 =
-\,(\phi_t)_*\big(\pounds_X\alpha_0\big)
 \,.
 \label{Push-forward-comp}
\end{align}
Equation \eqref{Push-forward-comp} is the push-forward Lie derivative formula. Note the opposite sign from the pull-back formula in \eqref{Pull-back-comp}. 
\vskip-28pt
\end{definition}
\begin{definition}\hskip-2pt
{\bf(Advected quantity)}\rm$\,$\\
An advected quantity is invariant along a flow trajectory. Thus, an advected quantity satisfies the pull-back relation
\[
\alpha_0(x_0) = \alpha_t(x_t) = (\phi_t^*\alpha_t)(x_0)
\,,\]
which implies the transport formula,
\begin{align}
0 =  \frac{d}{dt}\alpha_0(x_0)  =  \frac{d}{dt} (\phi_t^*\alpha_t)(x_0)
= \phi_t^* (\partial_t + \pounds_X)\alpha_t(x_0)
= (\partial_t + \pounds_X)\alpha_t(x_t)
\,,\label{advect-qty-PBdef}
\end{align}
where the vector field $X=\dot{\phi}_t\phi_t^{-1}$ generates the flow map $\phi_t$.

Equivalently, via the push-forward relation,
\[
\alpha_t (x_t) = (\alpha_0 \circ \phi_t^{-1})(x_t)
= ((\phi_t)_*\alpha_0) (x_t)
\,,\]
an advected quantity satisfies
\begin{align}
 \frac{d}{dt}\alpha_t(x_t)
 =
 \frac{d}{dt}(\phi_t)_*\alpha_0
 =
- (\pounds_X\alpha_t) (x_t)
 \,.
\label{advect-qty-PBdef}
\end{align}
\vskip-28pt
\end{definition}

\subsubsection{Pull-backs, push-forwards and Lie derivatives for GLM}

The GLM theory introduces a composition of maps, in which $\phi_{t,t/\eps} = \wt{g}_{t/\eps}\circ \ob{g}_t$ and whose pull-back satisfies the relation,
\[
\big(\wt{g}_{t/\eps}\circ \ob{g}_t \big)^* = \ob{g_t}^{\,*\,} \wt{g}_{t/\eps}^{\,*}\,.
\]
Advection by the composition of maps $\phi_{t,t/\eps} = \wt{g}_{t/\eps}\circ \ob{g}_t$ 
with vector fields $X:=\dot{\phi}_{t,t/\eps}\phi_{t,t/\eps}^{-1}$ and $\ob{X}:=\dot{\ob{g}}_t\ob{g}_t^{-1}$ satisfies 
the pull-back formula for the action of the composite transformation $$\phi_{t,t/\eps} = \wt{g}_{t/\eps}\circ \ob{g}_t$$ on a differential $k$-form or tensor field $\alpha$,%
\footnote{The notation $\ob{(\,\cdot\,)}$ and $\wt{(\,\cdot\,)}$ signifies time scales $t$ and $t/\eps$, respectively. Hence, we can drop subscripts as needed to simplify notation.}
\begin{align*}
 \frac{d}{dt}\big( ( \wt{g}\circ \ob{g})^* \alpha \big)
 =
( \wt{g}\circ \ob{g})^* \big(\p_t  \alpha+ \pounds_X\alpha\big)
 \,.
\end{align*}
Equivalently, the pull-back of the composition satisfies the relation
\begin{align*}
 \frac{d}{dt}\big( \ob{g}\,^* \wt{g}\,^* \alpha \big)
 =
\ob{g}\,^* \wt{g}\,^*\big(\p_t \alpha + \pounds_X\alpha\big)
 \,.
\end{align*}
Expanding out the time derivatives gives the following \emph{composite advective transport equation}
\[
0 = \big(\p_t + \mathcal{L}_X\big)\alpha
=
\wt{g}_*\ob{g}_* \frac{d}{dt} \big( \ob{g}\,^* \wt{g}\,^* \alpha \big) 
= \wt{g}_*\ob{g}_* \ob{g}\,^* \big( \p_t( \wt{g}\,^* \alpha) + \mathcal{L}_{\ob{X}}( \wt{g}\,^* \alpha)\big)
= \wt{g}_* \big( \p_t( \wt{g}\,^* \alpha) + \mathcal{L}_{\ob{X}}( \wt{g}\,^* \alpha)\big)
\,.\]
Recall that the pull-back, $\wt{g}^*$, is the inverse of the push-forward, $\wt{g}_*$. Hence, the pull-back of the previous formula by $\wt{g}^*$ implies the following version of the composite Lie transport formula, cf. \cite{GiVa2018},
\begin{align}
\wt{g}\,^*\Big( \big(\p_t + \mathcal{L}_X\big)\alpha\Big) = \big(\p_t  + \mathcal{L}_{\ob{X}}\big)( \wt{g}\,^* \alpha)
= 0\,.
\label{CompLieTrans-rel}
\end{align}


\subsection{GLM advective transport relations for Euler--Boussinesq
}\label{GLM-advection-sec}

For GLM, the smooth fast-time flow map on the manifold $M$ is taken to be $\wt{g}_{t/\eps}(M):=Id + \wt{\gamma}_{t/\eps}(M)$, where $\wt{\gamma}_{t/\eps}$ is a smooth invertible map parameterized by the fast time, $t/\eps$. This yields the familiar GLM fluctuation expression, $\wt{g}_{t/\eps}\bx = \bx + {\bm \xi}(\bx, t/\eps) = \bx^\xi$, when $M$ is taken to be $\mathbb{R}^3$. 
Consequently, formula \eqref{CompLieTrans-rel}  expands out in the GLM notation, to become
\begin{align*}
\Big(\big(\p_t + \mathcal{L}_X\big)\alpha\Big)(\bx + {\bm \xi}(\bx, t/\eps),t) 
&= \Big(\big(\p_t + \mathcal{L}_X\big)\alpha\Big)(\bx^\xi,t)
= \Big(\big(\p_t + \mathcal{L}_X\big)\alpha\Big)^\xi(\bx,t)
\\&= \big(\p_t  + \mathcal{L}_{\ob{X}}\big)( \wt{g}\,^* \alpha)
= \Big(\big(\p_t  + \mathcal{L}_{\ob{X}}\big)\alpha\Big)(\bx + {\bm \xi}(\bx, t/\eps),t)
\\&
= \big(\p_t  + \mathcal{L}_{\ob{X}}\big)\alpha(\bx^\xi,t)
= 0\,.
\end{align*}

Thus, the expansion of the composite advective Lie transport formula \eqref{CompLieTrans-rel} implies the following advective transport formula for a $k$ form $\alpha$, 
\begin{align}
\Big(\big(\p_t + \mathcal{L}_X\big)\alpha\Big)^\xi(\bx,t) 
&= 
\big(\p_t  + \mathcal{L}_{\ob{X}}\big) \alpha^\xi(\bx,t)
=
0\,.
\label{advect-rel1}
\end{align}
By a final transformation of variables, we will write the advection law \eqref{advect-rel1} as
\begin{align}
\big(\p_t  + \mathcal{L}_{\ob{X}}\big) \big( \widetilde{a}(\bx,t)\cdot de(\bx) \big)
=
0\,.
\label{advect-rel2}
\end{align}
This can be done by making the following chain rule calculation for the transformation of the tensor basis of $\alpha^\xi(\bx,t)$ in \eqref{advect-rel1},
\begin{align}
\alpha^\xi(\bx,t) =: a^\xi(\bx,t) \cdot de^\xi(\bx,t)
= \left(a^\xi(\bx,t) \cdot \frac{\p e^\xi(\bx)}{\p e(\bx)}\right)\cdot de(\bx)
=: \widetilde{a}(\bx,t)\cdot de(\bx) 
=: \wt{\alpha}(\bx,t)\,.
\label{advect-rel3}
\end{align}
Here,
$de(\bx)$ is the basis of the advected differential form or tensor, the quantity $\wt{a}(\bx,t)$ is its tensor coefficient in Eulerian coordinates and the centred dot denotes contraction of tensor indices. 

Equation \eqref{advect-rel3} implies that if $\alpha^\xi(\bx,t)$ is advected by $\bu^\xi$, then $\wt{\alpha}(\bx,t)$ will be advected by $\ob{\bf u}^L$. This is because the fluctuating quantity $\wt{\alpha}(\bx,t)$ defined above is merely a change of variables of $\alpha^\xi(\bx,t)$ from $\bx^\xi$ to $\bx$ via the chain rule. Moreover, the Eulerian mean of the relation \eqref{advect-rel3} yields
\begin{align}
 \ob{\left(a^\xi(\bx,t) \cdot \frac{\p e^\xi(\bx)}{\p e(\bx)}\right) } = \ob{\wt{\alpha}} = \wt{\alpha}\,.
\label{advect-rel4}
\end{align}
In taking this Eulerian mean, we keep in mind that $\mathbf{x}$ is an
average quantity, so the right hand side is {\it already} an average
quantity. Thus, $\wt{\alpha}$ satisfies $\ob{\wt{\alpha}}=\wt{\alpha}$ in \eqref{advect-rel4} and we note that
$\wt{\alpha} \ne \ob{\alpha}^L$, in general, except for the case that $\alpha^\xi$ is a scalar. 
The difference is that the tensor basis must be transformed to fixed Eulerian variables before applying the Eulerian time average, and a scalar function has no tensor basis. 

\begin{remark}[Road map for the remainder of the paper.]\rm
In principle, the fast-slow time-mean considerations underlying GLM described above could be generalized to the class of stochastic perturbations in \cite{Holm2015,HolmRT2018} whose analytical properties were examined in \cite{CrFlHo2018} by using the method of multi-time homogenisation \cite{GoMe2013a,GoMe2013b}  and by invoking the procedure for transition from a fast-slow description to the stochastic description for fluid dynamics developed in \cite{CoGoHo2017}. However, instead of launching into such an investigation by starting over from a stochastic viewpoint, we will build on the deterministic theory described in the Appendix to reach the point of introducing stochastic closures for the deterministic GLM description in Section \ref{SPDEs-sec}. 

In Section \ref{StochTrans-sec}, we will take advantage of the result of \cite{BdLHoLuTa2019} which proves the stochastic version of the pull-back formula \eqref{Pull-back-comp} for the Lie derivative with respect to a stochastic vector field. This result will allow us to introduce a class of stochastic closures of GLM in Section \ref{GLMclosures-sec}, each of which preserves its transport structure and fits into earlier work on stochastic fluid dynamics \cite{Holm2015,HolmRT2018,DrHo2019}. In Sections and \ref{Closure1a-subsec} and \ref{Closure1b-subsec}, we will suggest a simplified version of one of the closure models which we expect will be convenient in potential applications for analysis of GLM investigations of WCI elsewhere. 

The GM approach is adapted to the SALT framework in Section \ref{GMapproach-sec}.
Unlike the GLM model, which requires some development to cast it into the SALT framework, once the Gent-McWilliams (GM) approach is derived from a variational principle in Section \ref{GMrev-subsec}, it rather easily adapts to the SALT framework for uncertainty quantification in Section \ref{GM-SALT-subsec}.
\end{remark}


\subsection{GLM circulation transport}

As an example, we shall apply the composite Lie transport formula in \eqref{CompLieTrans-rel} 
to calculate the composite rate of change of the Kelvin circulation integral for the case
$\alpha = u(\bx,t)\cdot d\bx$, as follows
\begin{align}
\begin{split}
\frac{d}{dt}\oint_{g_{t,t/\eps}\gamma(\bx_0)} u(\bx,t)\cdot d\bx
&=
\oint_{\gamma(\bx_0)} \frac{d}{dt} \Big(g_{t,t/\eps}^*\big ( u(\bx,t)\cdot d\bx \big) \Big)
\\&=
\oint_{\gamma(\bx_0)} \frac{d}{dt} \Big( \ob{g}\,^* \wt{g}\,^* \big ( u(\bx,t)\cdot d\bx \big) \Big)
\\&=
\oint_{\gamma(\bx_0)} 
\ob{g}\,^* \wt{g}\,^* \Big( (\p_t  + \pounds_X  )  \big ( u(\bx,t)\cdot d\bx \big) \Big) 
\\&=
\oint_{\gamma(\ob{\bx})} 
\wt{g}\,^* \Big( (\p_t  + \pounds_X  )  \big ( u(\bx,t)\cdot d\bx \big) \Big) 
\\ \hbox{By \eqref{CompLieTrans-rel}}\ &=
\oint_{\gamma(\ob{\bx})}  (\p_t  + \pounds_{\ob{X} } ) 
 \Big( \wt{g}\,^* \big ( u(\bx,t)\cdot d\bx \big) \Big) 
\,.\end{split}
\label{circ-trans1}
\end{align}
Now, if $\wt{g}_{t/\eps}:=Id + \wt{\gamma}_{t/\eps}$, then $\wt{g}_{t/\eps}\bx = \bx + {\bm \xi}(\bx, t/\eps) = \bx^\xi$,
and the previous formula expands out in the GLM notation, to become
\begin{align}
\begin{split}
\frac{d}{dt}\oint_{g_{t,t/\eps}\gamma(\bx_0)} u(\bx,t)\cdot d\bx
&=
\oint_{\gamma(\ob{\bx})}  \big(\p_t  + \pounds_{\ob{X} } \big) \big(  u_i^\xi(\bx,t) J^i_j (\bx,t) \,dx^j \big)
\,,
\end{split}
\label{circ-trans2}
\end{align}
where $J^i_j $ is the GLM fluctuating Jacobian matrix
\begin{align}
J^i_j = \frac{\p x^{\xi \, i}} {\p x^j } = 
\Big(\delta^i_j + \frac{\partial \xi^i}{\partial x^j} \Big) \,.
\end{align}
Consequently, the 1-form in the integrand of \eqref{circ-trans2} becomes, upon assuming that
$\ob{X} := \dot{\ob{g}}_t\ob{g}_t^{-1}=\ob{u}^L$,
\begin{align}
\wt{u}_i \,dx^i
:=
u_i^\xi(\bx,t) J^i_j (\bx,t) \,dx^j 
= \big(\ob{u}^L_i + u^\ell_i  \big) \Big(\delta^i_j + \frac{\partial \xi^i}{\partial x^j} \Big) \,dx^j 
\end{align}
whose Eulerian time average is 
\begin{align}
\ob{\wt{u}}_i \,dx^i = \Big(\ob{u}^L_i + \ob{u^\ell_j \,\partial_i \,\xi^j} \,\Big)\cdot dx^i
\,.
\end{align}
Thus, we may conclude the following formula for the rate of change 
of the fast-time average of the Kelvin circulation integral,
\begin{align}
\frac{d}{dt} \ob{ \oint_{\gamma(\bx^\xi)} \bu^\xi(\bx,t)\cdot d\bx^\xi }
=
\oint_{\gamma(\ob{\bx})}  \big(\p_t  + \pounds_{ \ob{u}^L } \big) 
\Big( \big(\ob{\bu}^L_i + \ob{u^\ell_j \,\nabla \,\xi^j} \,\big)\cdot d\bx \Big)
\,.
\label{circ-trans-pseudomomap}
\end{align}
As we shall see, formula \eqref{circ-trans-pseudomomap} is the basis for the definition of the \textit{pseudomomentum} in the GLM theory. 


\subsection*{GLM scalar advection relations}

Now that we have explained how the pull-back formula \eqref{CompLieTrans-rel} 
implies the Lie derivative description of 
advective transport for GLM, we may return to the classic notation of GLM to discuss examples.

At fixed position $\mathbf{x}$ the GLM velocity decomposition $\mathbf{u}^\xi =
\overline{\mathbf{u}}^L + \mathbf{u}^\ell$ is the sum of the
Lagrangian mean velocity $\overline{\mathbf{u}}^L$ and the Lagrangian
disturbance velocity $\mathbf{u}^\ell$. Thus, 
\[
\mathbf{u}^\xi = \frac{D^L\mathbf{x}^\xi}{Dt} 
\] 

and for any \emph{scalar} field $\chi(\mathbf{x},t)$ one has, 
\begin{equation}
\Big(\frac{D\chi}{Dt}\Big)^\xi
=
\frac{D^L}{Dt}(\chi^\xi)
\,.\nonumber
\end{equation}
Because $\overline{\mathbf{u}}^L$ appearing in the advection operator 
${D^L}/{Dt}=\partial_t+\overline{\mathbf{u}}^L\cdot\nabla$ is a mean
quantity, one then finds, as expected, that the Lagrangian mean
$\overline{(\,\cdot\,)}^L$ commutes with the original material time derivative
$D/Dt$ for a scalar function. That is,
\begin{equation}
\overline{\Big(\frac{D\chi}{Dt}\Big)}^L
=
\frac{D^L}{Dt}(\overline{\chi}^L)
\,,\quad\hbox{and}\quad
\Big(\frac{D\chi}{Dt}\Big)^\ell
=
\frac{D^L}{Dt}\chi^\ell
\,,\nonumber
\end{equation}
where $\chi^\ell=\chi^\xi-\overline{\chi}^L$ is the Lagrangian
disturbance of $\chi$ satisfying $\overline{\chi^\ell}=0$. 
Hence, one finds several equivalence relations for scalars, cf. formulas \eqref{advect-rel2} 
and \eqref{advect-rel3},
\begin{equation}
\Big(\frac{D\chi}{Dt}\Big)^\xi
=
\frac{D^L}{Dt}(\chi^\xi)
=
\overline{\Big(\frac{D\chi}{Dt}\Big)}^L + \Big(\frac{D\chi}{Dt}\Big)^\ell
=
\frac{D^L}{Dt}(\overline{\chi}^L) + \frac{D^L}{Dt}\chi^\ell
\,.
\label{Ddt-equiv-rels}
\end{equation}
For
example, in the Euler-Boussinesq stratified incompressible flow, consider 
the buoyancy $b=(\rho_{ref}-\rho)/\rho_{ref}$ relative to a reference density $\rho_{ref}$. 
In this case, the buoyancy $b$ is advected as a scalar function. That is, it satisfies $Db/Dt=0$ and,
by the relations \eqref{Ddt-equiv-rels}, the average yields $D^L\overline{b}^L/Dt=0$, as well. Hence,
$b^\xi=\overline{b}^L$ follows, by integration of
$D^L(\overline{b}^L-b^\xi)/Dt=0$ along mean trajectories and invertibility
of the map $\mathbf{x}\to\mathbf{x}+\xi(\mathbf{x},t)$. 

\begin{remark}\rm
Of course, this identification is also obvious
physically for scalars, since the Lagrangian mean $\overline{b}^L$ and the current
value $b^\xi$ refer to the {\it same} Lagrangian fluid label. 
That is, we initialize with $\xi(\mathbf{x}_0,0)=0$, for a Lagrangian
coordinate $\mathbf{x}_0=\mathbf{x}(\mathbf{x}_0,0)$.
\end{remark}


\subsection*{Mass conservation: the GLM continuity equation}

The instantaneous mass conservation relation $D^\xi(x,t)\,d^3x^\xi = D(x_0)d^3x_0$ transforms into current Eulerian coordinates as follows, cf. equation \eqref{advect-rel3},
{
\begin{equation}
D^\xi\,d^3x^\xi 
=
D^\xi{\cal J} \,d^3x
=:
\widetilde{D}\,d^3x
= D(x_0)d^3x_0
\,,
\label{Dtilde-def}
\end{equation}
where one defines the Jacobian,
\begin{equation}
{\cal J}
=
\det\big(\nabla_\mathbf{x}(\mathbf{x}+\xi)\big)
=
\det\Big(\delta^i_j + \frac{\partial \xi^i}{\partial x^j} \Big)
,\quad\hbox{and}\quad
\widetilde{D} := D^\xi{\cal J} 
\,.
\label{DensityConstraint}
\end{equation}
}
As in the previous section, in taking the Eulerian mean of the relation $D^\xi{\cal J} \,d^3x =
\widetilde{D}\,d^3x$, we keep in mind that $\mathbf{x}$ is an
average quantity, so the right hand side is {\it already} an average
quantity. Thus, $\widetilde{D}=D^\xi{\cal J}$ satisfies
$\overline{\widetilde{D}}=\widetilde{D}$ and we note that
$\widetilde{D}\ne\overline{D}^L$, in general. 

The mean mass conservation relation for advection, $\wt{D}(x,t)d^3x=D(x_0)d^3x_0$, then implies the continuity
equation for $\widetilde{D}$,
\begin{equation}
\big( \p_t  + \mathcal{L}_{\ob{\mathbf{u}}^L}\big) \big( \wt{D} \,d^3x \big)
=
0\,,
\quad\Longrightarrow\quad
\partial_t\,\wt{D}
+
{\rm div}\wt{D}\ob{\mathbf{u}}^L
=
0
\,,
\label{contin-eqn}
\end{equation}
upon recalling that $\overline{\mathbf{u}}^L$ is the
velocity tangent to the mean Lagrangian position $\mathbf{x}$.
Consequently, the Lagrangian mean $\overline{D^\xi}=\overline{D}^L$ is not the density advected in the GLM theory.
Rather, it is the average density, $\overline{D^\xi{\cal J}}=\widetilde{D}$. As discussed in the previous section, except for scalars such as the buoyancy, $b$, this observation applies to all advected quantities. That is, the basis of any differential form or tensor field evolves under advection by the flow map, as well as its instantaneous coefficient.

\begin{remark}\rm
For a fluid with {\bf constant density}, $D^\xi=1$,
the GLM theory gives
\begin{equation}\label{GLM-density}
\widetilde{D}
=
\overline{ D^\xi{\cal J} }
=
\overline{ \det\big(\nabla_\mathbf{x}(\mathbf{x}+\xi)\big) }
=
1 - \tfrac{1}{2}
\big(\,\overline{ \xi^k\xi^\ell}\,\big)_{\!\!,\,k\ell}
+
O(|\xi|^3)
\,.\nonumber
\end{equation}
Hence, for constant instanteous density, the Lagrangian mean velocity
$\overline{\mathbf{u}}^L$ has an order $O(|\xi|^2)$ divergence, 
\[
{\rm div}\,\overline{\mathbf{u}}^L
=
-\,\frac{1}{\widetilde{D}}
\frac{D^L\widetilde{D}}{Dt}
=
\frac{1}{2}\frac{D^L}{Dt}
\big(\,\overline{ \xi^k\xi^\ell}\,\big)_{\!\!,\,k\ell}
+
O(|\xi|^3)
\,,\]
as shown in Andrews \& McIntyre [1978a] \cite{AM1978a}. 

\end{remark}

\section{Stochastic closures for the GLM equations}
\label{SPDEs-sec}

\subsection{Stochastic transport via the Kunita--It\^o--Wentzell formula}
\label{StochTrans-sec}

The remainder of this section will introduce stochastic closure schemes for the GLM and Gent--McWilliams (GM) models of mesoscale and sub-mesoscale transport. The key ingredient for these stochastic closure schemes will be the generalisation to stochastic processes of the pull-back formula for the Lie derivative in equation \eqref{Pull-back-comp}. This stochastic generalisation is given by
\begin{align}
 \frac{d}{dt}(\phi_t^*{K})
 =
 \phi_t^*\big(\p_t{K} + \pounds_{X_t}{K}\big)
 \,,
\label{Pull-back-comp-stoch}
\end{align}
where the time-dependent vector field $X_t\in \mathfrak{X}(M)$ generates the flow map $\phi_t$ via $\dot{\phi}_t = X_t \circ \phi_t$ and ${K} \in \Lambda^k(M)$ is a spatially smooth $k$-form on a manifold $M$. 

The corresponding pull-back formula for $k$-forms which are spatially smooth and stochastic in time is proven in \cite{BdLHoLuTa2019}. Namely,  in the standard  differential notation for stochastic flows, we have
\begin{align}
\diff \,(\phi_t^* {K})(t,x) = \phi_t^*\left(\diff {K} + \mathcal L_{\diff x_t} {K} \right)(t,x),
\end{align}
where $\diff x_t$ is the stochastic spatially smooth vector field defined by 
\begin{align}
\phi_t^*\diff x_t(x) = 
\diff x_t(\phi_t(x)) = b(t,\phi_t(x)) \diff t + \sum_{i=1}^N \xi_i(t,\phi_t(x)) \circ \diff B_t^i,
\end{align}
which generates the stochastic flow $\phi_t$ in equation \eqref{Pull-back-comp-stoch}.

Equation \eqref{Pull-back-comp-stoch} is the Kunita-It\^o-Wentzell formula \cite{Kun81,Kun84,Kun97} which determines the evolution of a $k$-form-valued stochastic process $\phi_t^*{K}$. This result generalises the classic formula for a stochastic scalar function by allowing $K$ to be any smooth-in-space, stochastic-in-time $k$-form on $\mathbb{R}^n$. Omitting the technical regularity assumptions provided in the more detailed statement of the theorem in \cite{BdLHoLuTa2019} for our purposes here, we now state a simplified version of the main theorem proved in that paper, as follows.

\begin{theorem}[Kunita-It\^o-Wentzell formula for $k$-forms, simplified version]\rm
Consider a spatially smooth $k$-form $K(t,x)$ which is a semimartingale in time
\begin{align} \label{spde-compact}
\diff K(t,x) = G(t,x) \diff t + \sum_{i=1}^M H_i(t,x) \circ \diff W_t^i,
\end{align}
where $W_t^i$ are i.i.d. Brownian motions. Let $\phi_t$ be a sufficiently smooth flow satisfying the SDE 
\begin{align}
\diff \phi_{t}(x) = X(t,\phi_{t}(x)) \diff t + \sum_{i=1}^N \zeta_i(t,\phi_{t}(x)) \circ \diff B_t^i
\,,
\end{align}
in which  $B_t^i$ are i.i.d. Brownian motions. Then the pull-back $\phi_t^* K$ satisfies the formula
\begin{align}
\begin{split}
\diff \,(\phi_t^* K)(t,x) = &\phi_t^* G(t,x) \diff t + \sum_{i=1}^M \phi_t^* H_i(t,x) \circ \diff W_t^i \\
&+  \phi_t^*\mathcal L_X K(t,x) \diff t  + \sum_{i=1}^N \phi_t^* \mathcal L_{\zeta_i} K(t,x) \circ \diff B_t^i
\,.
\label{KIWkformsimplified}
\end{split}
\end{align}
\end{theorem}
Formulas \eqref{spde-compact} and \eqref{KIWkformsimplified} are compact forms of the equations derived in \cite{BdLHoLuTa2019}, where these equations are written in integral notation to make the stochastic processes more explicit. However, the compact differential stochastic notation used here will suffice to explain the main ideas in the next section. For more details and proofs, see \cite{BdLHoLuTa2019}.


\subsection{Stochastic closures for GLM approximation of the Euler--Boussinesq equations}
\label{GLMclosures-sec}

So far, the WCI system in the GLM equation sets \eqref{Waves-SD-LPmatrix} and \eqref{Fluid-SD-LPbrkt} has not been closed. This is because the mean fluctuation quantities comprising the kinematic pressure $\ob{\pi^\ell}$ in \eqref{Pi-tot} and the relative group velocity $\ob{\bf v}_G$ in \eqref{group-vel-def} have not yet been parameterized. In this section, following Holm \cite{Holm2015} and Gay-Balmaz and Holm \cite{FGB-Ho2018}, we consider two different classes of closure options for modelling these unknown quantities stochastically. Simply put, the two different classes of closure are either (1) data-driven, or (2) model-driven. In more detail, the options are: (1) apply prescribed noise which has been calibrated from observations and simulations, or (2) postulate a theoretical model for the dynamics of the noise amplitude depending on advected state variables, such as the buoyancy. In either case, the result would provide an estimate of the uncertainty in the model computations, which in turn would provide opportunities for reduction of uncertainty by using data assimilation. 

\subsection{Stochastic GLM Closure \#1a}\label{Closure1a-subsec}

A very interesting approximation of the kinematic fluctuation pressure is discussed in \cite{AM1978b}; namely,
\begin{align}
-\,\ob{\pi^\ell}
=
\frac12 \ob{ | \bu^\ell |^2} +  \overline{ \bu^\ell\cdot\bR^\ell} \approx \ob{ p^\xi_{,j}K^j_i\xi^i  }\,.
  \label{Fluct-pressure}
\end{align}
Both this approximation and the relative group velocity $\ob{v}_G^j= \overline{(p^\xi K^j_i \,\p_\phi\xi^i)}$ in \eqref{group-vel-def} involve the time mean correlations among the fluctuation displacements $ \xi^i$ and the corresponding  fluctuating pressure $p^\xi$. 

This observation suggests that one could close the WCI system by introducing a stochastic parameterization of these undetermined time mean correlations among the fluctuating degrees of freedom appropriate to the variable over which one is averaging. For example, the stochastic parameterization could comprise a pair of Stratonovich stochastic process,
\[
\overline{\boldsymbol{v}}_G \to {\rm d}\overline{\boldsymbol{v}}_G = {\bm \zeta}(\bx)\circ dW_t
\,,\quad\hbox{and}\quad
\Pi_{tot}  \to {\rm d}\Pi_{tot} = \overline{\Pi}^Ldt + \pi(\bx) \circ dW_t
\,.
\]
In turn, this idea suggests is a new type of Hamiltonian stochastic closure which has been studied recently for fluid dynamics in \cite{Holm2015,HolmRT2018,CoGoHo2017,CrFlHo2018,CoCrHoPaSh2018a,CoCrHoPaSh2018b}. It amounts to changing the GLM Hamiltonian in equation \eqref{Hbar-det} into the following stochastic process,
\begin{align}
\begin{split}
{\rm d}\Hbar (\ob{\bf m}, N, \pb, \Dt, \bb^l; \omega, \bk, \overline{\bm v}_G)
&= \int \bigg[
\frac{1}{2\widetilde{D}}\big| \mb + \pb - \Dt\Rb^L  \big|^2 
+ N(\omega - \bk \cdot \ob{\bu}^L)
\\&\hspace{1cm}
+ 
\Dt \Big( \pbL + gz\bb^L + \PhibL(\bx) \Big)
\,\bigg]\,d^3x \, dt
\\&\hspace{2mm}
+ 
\int \bigg[ (\pb - N\bk )\cdot \zeta(\bx)
- \Dt \pi(\bx) \,\bigg]\,d^3x \,\circ dW_t
\,.
\end{split}
\label{Hbar-stoch}
\end{align}

\paragraph{Hamiltonian properties.} The stochastic GLM Euler--Boussinesq equations may be expressed in Hamiltonian Lie--Poisson matrix operator form as follows, in which the dynamics of the wave variables $\pb$ and $N$ acquires a stochastic component of its \emph{transport velocity}, as
\begin{align}
{\rm d}\!
\begin{bmatrix}\,
\ob{p}_j \\  \\ N 
\end{bmatrix}
= - 
   \begin{bmatrix}
   \ob{p}_k\partial_j + \partial_k \ob{p}_j &  N\partial_j 
   \\ \\
   \partial_k N & 0 &  
   \end{bmatrix}
   \begin{bmatrix}
{\delta ({\rm d}\Hbar)/\delta \ob{p}_k} = \ob{u}^{L\,k}dt  + \zeta^k(\bx)\circ dW_t \\ \\
{\delta ({\rm d}\Hbar)/\delta N} =  \omega dt - k_i (\ob{u}^{L\,i}dt + \zeta^i(\bx)\circ dW_t ) \\
\end{bmatrix}
\,.
  \label{Waves-SD-LPbrkt-SALT}
\end{align}

The dynamics of the material variables $\ob{m}_j, \Dt$ and $\bb^L$ acquires a stochastic component of its \emph{pressure force}, as
\begin{align}
{\rm d}\!
\begin{bmatrix}\,
\ob{m}_j\\ \Dt \\ \ \bb^L
\end{bmatrix}
= - 
   \begin{bmatrix}
   \ob{m}_k\partial_j + \partial_k \ob{m}_j &
   \Dt\partial_j &
   - \,\bb^L_{,j}  &
   \\
   \partial_k \Dt & 0 & 0 & 
   \\
   \bb^L_{,k} & 0 & 0 &  
   \end{bmatrix}
   \begin{bmatrix}
{\delta \Hbar/\delta \ob{m}_k} = \ob{u}^{L k} \,dt \\
{\delta \Hbar/\delta \Dt} =  \Big( \pbL + gz\bb^L + \PhibL(\bx) \Big)dt 
- \pi(\bx) \circ dW_t  \\
{\delta \Hbar/\delta \bb^L} = \widetilde{D}\,gz  \,dt \\
\end{bmatrix}
\,.
  \label{Fluid-SD-LPbrkt-SALT}
\end{align}
This stochastic pressure force does not affect the fluid circulation in Kelvin's theorem in equation \eqref{GLM-EB-Kelvin-thm}. 

In the stochastic representation of fluctuating wave effects in the GLM picture, the stochastic pressure fluctuations in \eqref{Fluid-SD-LPbrkt-SALT} might arguably be dropped because they cause no circulation. In that case, the stochasticity of the GLM group velocity in \eqref{Waves-SD-LPbrkt-SALT} would coincide with the existing theory of Stochastic Advection by Lie Transport (SALT) \cite{Holm2015,HolmRT2018,CoGoHo2017,CrFlHo2018,CoCrHoPaSh2018a,CoCrHoPaSh2018b} which  introduces the same type of Hamiltonian stochastic transport into the material fluid evolution. 

\subsection{Stochastic GLM Closure \#1b}\label{Closure1b-subsec}

Perhaps the straightest way toward the introduction of stochastic effects in WCI for use in uncertainty quantification and future data assimilation would be to consolidate the stochasticity of the GLM group velocity with the known SALT approach of adding a stochastic vector field to the Lagrangian mean transport drift velocity, $\ob{\bf u}^L\,dt$, rather than proliferating the possible sources of uncertainty by making the GLM group velocity independently stochastic. In the SALT procedure, one takes the noise to be a $\sum_{a=1}^N{\bm \zeta}_a(\bx)\circ dW^a_t$ in which each stochastic spatial `mode' ${\bm \zeta}_a(\bx)$ is associated to a different Brownian motion $dW^a_t$ and must be calibrated, for example, by comparison of high resolution data from either observation or computational simulation. To simplify the notation in this section, we neglect the option to include modal spatial structure in the noise by ignoring the sum over indices for the individual Brownian motions. 

In the closure strategy \#1b, both wave and fluid dynamics would acquire the \emph{same} fluctuating component in the GLM \emph{transport velocity}, as
\begin{align}
{\rm d}\!
\begin{bmatrix}\,
\ob{p}_j \\  \\ N 
\end{bmatrix}
= - 
   \begin{bmatrix}
   \ob{p}_k\partial_j + \partial_k \ob{p}_j &  N\partial_j 
   \\ \\
   \partial_k N & 0 &  
   \end{bmatrix}
   \begin{bmatrix}
{\delta ({\rm d}\Hbar)/\delta \ob{p}_k} = \overline{u}^{L\,k}dt  + \zeta^k(\bx)\circ dW_t \\ \\
{\delta ({\rm d}\Hbar)/\delta N} =  \omega dt - k_i (\ob{u}^{L\,i}dt + \zeta^i(\bx)\circ dW_t ) \\
\end{bmatrix}
\,,
  \label{Waves-SD-LPbrkt-stoch}
\end{align}
for the waves, and 
\begin{align}
{\rm d}\!
\begin{bmatrix}\,
\ob{m}_j\\ \Dt \\ \ \bb^L
\end{bmatrix}
= - 
   \begin{bmatrix}
   \ob{m}_k\partial_j + \partial_k \ob{m}_j &
   \Dt\partial_j &
   - \,\bb^L_{,j}  &
   \\
   \partial_k \Dt & 0 & 0 & 
   \\
   \bb^L_{,k} & 0 & 0 &  
   \end{bmatrix}
   \begin{bmatrix}
{\delta \Hbar/\delta \ob{m}_k} = \ob{u}^{L\,k}dt  + \zeta^k(\bx)\circ dW_t \\
{\delta \Hbar/\delta \Dt} =  \Big( \pbL + gz\bb^L + \PhibL(\bx) \Big)dt  \\
{\delta \Hbar/\delta \bb^L} = \widetilde{D}\,gz  \,dt \\
\end{bmatrix}
\,,
  \label{Fluid-SD-LPbrkt-stoch}
\end{align}
for the fluid, where we recall that 
$\ob{\mathbf{m}}+\ob{\mathbf{p}} =  \Dt(\ob{\mathbf{u}}^L + \ob{\mathbf{R}}^L)$
and $\ob{\mathbf{p}}= \Dt\ob{\mathbf{v}}$.

This means the GLM Kelvin circulation theorem for  Boussinesq incompressible flow
in equation \eqref{GLM-comp-Kel-thm} will become
\begin{equation}\label{GLM-comp-Kel-thm-stoch}
{\rm d}\oint_{c({\rm d}\bx_t)}
\Dt^{-1} \,\ob{\mathbf{m}} \cdot d\mathbf{x}
=
{\rm d}\oint_{c({\rm d}\bx_t)}
\Big( \ob{\mathbf{u}}^L + \ob{\mathbf{R}}^L - \ob{\mathbf{v}}
\Big)
\cdot
d\mathbf{x}
= -g\,
\oint_{c({\rm d}\bx_t) }
\overline{b}^L \,dz
\,,
\end{equation}
in which the material loop moves along stochastic Lagrangian trajectories given by 
the characteristics of the following stochastic vector field 
\begin{equation}\label{GLM-comp-Kel-thm-stochpath}
{\rm d}\bx_t = \ob{\bu}^L(\bx_t,t)dt  + \sum_{a=1}^N  {\bm\zeta}_a(\bx_t)\circ dW^a_t 
\,.
\end{equation}
Adding the stochastic vector field into \eqref{GLM-comp-Kel-thm-stochpath} amounts to modifying the 
final term in the stochastic GLM Hamiltonian in equation \eqref{Hbar-stoch}, as follows,
\begin{align}
\begin{split}
{\rm d}\Hbar (\ob{\bf m}, N, \pb, \Dt, \bb^l; \omega, \bk, \overline{\bm v}_G)
&= \int \bigg[
\frac{1}{2\widetilde{D}}\big| \mb + \pb - \Dt\Rb^L  \big|^2 
+ N(\omega - \bk \cdot \ob{\bu}^L)
\\&\hspace{1cm}
+ 
\Dt \Big( \pbL + gz\bb^L + \PhibL(\bx) \Big)
\,\bigg]\,d^3x \, dt
\\&\hspace{2mm}
+ 
\int \bigg[ (\ob{\bf m} + \pb - N\bk )\cdot  \sum_{a=1}^N {\bm\zeta}_a(\bx)\,\bigg]\,d^3x \,\circ dW^a_t
\,.
\end{split}
\label{Hbar-stoch-SALT}
\end{align}
%

\begin{remark}\rm
In the class of closures \#1a and \#1b, with prescribed noise, it still remains to determine the set of vectors $\{ {\bm\zeta}_a(\bx_t)\}$ in the stochastic part of the Lagrangian trajectory given by ${\rm d}\bx_t$ in equation \eqref{GLM-comp-Kel-thm-stochpath}. For this, it may be advisable to model the effects of wave fluctuations in the GLM equations \eqref{GLM-comp-Kel-thm-stoch} and \eqref{GLM-comp-Kel-thm-stochpath} the same way as for any other high frequency transport effect in the SALT modelling approach of \cite{CoGoHo2017, Holm2015, HolmRT2018}.  This approach would also simplify the calibration procedure for the correlation eigenvectors in ${\bm \zeta}(\bx)\circ dW_t$, which is required in the application of SALT, because it would consolidate the stochastic effects of the wave transport with those of the material transport. Distinguishing between these two types of stochastic effects in the total transport by using observation data might be problematic, to say the least. For recent developments using the SALT approach to material transport and the description of the use of data assimilation in determining the stochastic amplitudes in two-dimensional flows, see \cite{CoCrHoPaSh2018a, CoCrHoPaSh2018b}.
\end{remark}

\section{The Gent-McWilliams (GM) approach}\label{GMapproach-sec}

\subsection{Brief review of the deterministic GM approach}\label{GMrev-subsec}

The SALT approach  could be regarded as a data-driven stochastic version of the Gent-McWilliams (GM) parameterization of subgrid-scale transport  \cite{Gent2011,GM1990,GM1996}, which is commonly used in both ocean and atmospheric sciences. In a landmark paper \cite{GM1990}, Gent and McWilliams modified passive tracer advection by adding a term meant to model eddy transport. The GM term introduced an anisotropic model of fluid transport which depends on the local gradients of the buoyancy. This term is still used today in the large majority of ocean models.  Since the wave component of the GLM theory fundamentally depends on buoyancy, one can imagine that the two approaches could interact with each other  synergistically. For this purpose, we will first briefly review the GM approach in the present notation. Then, we will discuss how Model 3 in \cite{FGB-Ho2018} enables one to build on the GM approach and construct a stochastic closure of the motion equation in which the spatial correlations of the stochasticity depend on the quantities advected by the flow.  

\paragraph{Geometry of the GM approach.}
Let $u(x,t)$ be a fluid velocity variable, and let $a(x,t)$ be an advected variable.
The GM approach begins by introducing a modified transport equation for advection, as
\begin{align}
\p_t a + \L_U a = 0
\,\quad\hbox{with}\quad 
U = u +u^*(a)\,,
\label{aux-eqn}
\end{align}
where $\L_U a$ is the Lie derivative of the advected variable $a$ with respect to the vector field $U$, and the GM model \emph{bolus velocity} $u^*(a,a_{,j},a_{,jk})$ is a prescribed vector function of $a$ and its first two spatial derivatives. In particular, the GM model takes the advected quantity $a$ to be the buoyancy, $b$, which is a scalar function. 

To find the effect on the motion equation of modifying the advection law in \eqref{aux-eqn} one may use a Lagrange multiplier to constrain Hamilton's variational principle for ideal fluids $\de S=0$ with $S=  \int \ell (\bu, a)\,dt$ to satisfy the modified auxiliary advection equation \eqref{aux-eqn}. Before taking variations, one defines  the following useful notational constructs. 
\begin{enumerate}
\item
Define $V(M)$ a vector space defined on the domain of flow, $M$,
as well as $ \mathfrak{X}(M)$ the space of smooth vector fields defined on $M$.  

\item
Define real, non-degenerate $L^2$ pairings between the spaces $V(M)$ and $ \mathfrak{X}(M)$ 
with their dual spaces, $V^*(M)$ and $ \mathfrak{X}^*(M)$
\[
\scp{\,\cdot\, }{\,\cdot\,}_V: V^*\times V\to\mathbb{R}
,\quad 
\scp{\,\cdot\, }{\,\cdot\,}_\mathfrak{X}: \mathfrak{X}^*\times \mathfrak{X}\to\mathbb{R}
. 
\]
\item
Define the diamond operator $(\diamond)$ in terms of these two pairings and the Lie derivative, as
\[
\scp{\pi}{\L_{\de u}a}_V = - \scp{\pi\diamond a}{\de u}_\mathfrak{X}
\,,\]
for $a\in V$, $\pi\in V^*$ and $\de u \in \mathfrak{X}(M)$. Thus, $\L_{\de u}a$ is the Lie derivative of the advected quantity $a$ in the direction of the velocity variation $\de u$.
\end{enumerate}

To determine the effect on the motion equation of modifying the advection law in \eqref{aux-eqn}, we apply the auxiliary equation \eqref{aux-eqn} as a constraint in Hamilton's principle for ideal fluids. Namely,
we constrain Hamilton's variational principle $\de S=0$ with $S=  \int \ell (\bu, a)\,dt$ 
to advect the quantity $a$ by a total $U = u + u^*(a)$,  
by pairing equation \eqref{aux-eqn} with a Lagrange multiplier, $\pi$. Thus, we set
\begin{align}
0 = \de S &= \de \int \left[\ell (\bu, a) + \Scp{\pi}{\p_t a + \L_U a }_V \right]\,dt\,.
\end{align}
We then take variations to find,\footnote{When performing integration by parts in the variational principle, one assumes homogeneous boundary conditions.}
\begin{align}
\begin{split}
\de u: & \quad  \frac{\de \ell}{\de u} = \pi\diamond a
\,,\\
\de \pi: & \quad \p_t a = -\, \L_U a 
\,,\\
\de a: & \quad \p_t \pi = \L^T_{U}\pi + \frac{\de \ell}{\de a} - \gamma
\,\quad\hbox{with}\quad 
\gamma := \left( \frac{\de u^*}{\de a} \cdot \frac{\de \ell}{\de u} \right)
,\end{split}
\label{tally-var}
\end{align}
and manipulate further to obtain the following Euler--Poincar\'e equations \cite{Holm2015,HoMa2019,HMR1998},
\begin{align}
\begin{split}
\p_t  \frac{\de \ell}{\de u} + \L_U  \frac{\de \ell}{\de u}  
&=
\Big(\frac{\de \ell}{\de a} - \gamma \Big)\diamond a 
\,\quad\hbox{with}\quad 
\gamma := \left( \frac{\de u^*}{\de a} \cdot \frac{\de \ell}{\de u} \right)
\,,\\
\p_t a + \L_U a &= 0
\,\quad\hbox{with}\quad 
U = u +u^*(a)
\,.\end{split}
\label{gen-eqns}
\end{align}
The variation $\de u^*/\de a$ of the prescribed bolus velocity $u^*(a)$ with respect to the advected variable $a$ results in a \emph{differential operator} in the $\gamma$-term, which arises from integration by parts in the $\de a$-variations, contracted with  the variation $\de \ell/\de u^i $, for example, as
\begin{align}
\left( \frac{\de u^* (a, a_{,j}, a_{,jk})}{\de a} \right) \cdot \frac{\de \ell}{\de u}   
:=  \frac{\p u^{*\,i} }{\p a} \frac{\de \ell}{\de u^i}  - \,\p_j \left( \frac{\p u^{*\,i} }{\p a_{,j}} \frac{\de \ell}{\de u^i} \right)
+ \p^2_{jk}  \left(\frac{\p u^{*\,i} }{\p a_{,kj}}  \frac{\de \ell}{\de u^i} \right)
\,.
\label{a-var-eqn}
\end{align}
The GM choice for $\bu^*(b)$ in terms of the advected buoyancy $b(\bx,t)$ is linearly proportional to the local isopycnal slope ${\bf s}=-(\nabla_H b)/ b_z$, namely,
\begin{align}
\bu^*(b, b_{,j}, b_{,jk}) = {\rm curl} (\, {\bf \widehat{z}} \times \kappa {\bf s}\,)
= -  {\bf \widehat{z}}\cdot \nabla (\kappa {\bf s}) = \p_z\left(\frac{\kappa \nabla_H b}{b_z}\right)
,\label{GM-BV}
\end{align}
where $\nabla_H$ is the horizontal gradient. Note that ${\rm div}\bu^*(b)=0$. Consequently, the pressure for is determined by taking the divergence of the motion equation for incompressible flow, as usual. Upon denoting $\de \ell/\de \bu = {\bf m} $, one evaluates the operator in equation \eqref{a-var-eqn} for constant scalar $\kappa$ as
\begin{align}
\frac{\de \bu^* (b)}{\de b}\cdot {\bf m}
= 
-\,\p_z\bigg(\frac{\kappa}{b_z^2}\nabla b\cdot \p_z{\bf m}\bigg)
+\nabla\cdot\bigg(\frac{\kappa}{b_z}\p_z{\bf m}\bigg)
\,.
\label{Bolus-op}
\end{align}

The scalar advection $\bU\cdot\nabla \frac{\de \ell}{\de u}$ part of the momentum transport $ \L_U  \frac{\de \ell}{\de u} $ in equation \eqref{gen-eqns} appeared in equations (8) and (9) of  \cite{GM1996}, where its magnitude was estimated as order the Rossby number, $\eps$, so that $U = u +\eps u^*(b)$. Thus, according to  \cite{Gent2011}, this term would make little difference in computational simulations at non-eddy-resolving resolution; so, it has never been implemented in an ocean climate computation. However, it could make a difference in ageostrophic situations, where finer resolution is required.  See, e.g., \cite{Fox-etal2008,HCDF-K-ocean2017,GrKl2019} for the latest investigations of this point. 

\begin{remark}[Outlook]\rm
We are starting with the GM modification in the transport velocity and deriving its consequences via the variational principle for ideal fluid dynamics. The resulting variational Gent-McWilliams model (VGM) will differ from the original GM equations \cite{GM1990,GM1996} in its momentum balance, energetics, Kelvin's circulation theorem and potential vorticity conservation on fluid particles. Then we will introduce stochastic transport in the VGM setting.

\end{remark}

\subsubsection{Example: Euler--Boussinesq equations}

For $a=(b,D)\in V\times V^*$ for scalar buoyancy $b\in \Lambda^0$ and mass density $D\in \Lambda^3$ in 3D, the diamond operations in these equations may be expressed as follows
\begin{align}
\begin{split}
\scp{\gamma\diamond b}{\eta} &= \int \gamma (-{\bm \eta}\cdot \nabla b)\,d^3x
= - \int (\gamma \nabla b) \cdot {\bm \eta}\,d^3x 
= - \scp{\gamma d b\otimes d^3x}{\eta}
,\\
\Scp{\frac{\de \ell}{\de D}\diamond D }{\eta} &= - \int \frac{\de \ell}{\de D}\, {\rm div} (D{\bm \eta} )\,d^3x
=  \int D \nabla \frac{\de \ell}{\de D} \cdot {\bm \eta} \,d^3x
= \Scp{D d \frac{\de \ell}{\de D} \otimes d^3x }{\eta}
\,.\end{split}
\label{diamond-eqns}
\end{align}

The Lagrangian in Hamilton's principle for the Euler--Boussinesq equations is 
\begin{align}
\ell(\bu,D,b) = \int D \left( \frac12 |\bu|^2 + \bu\cdot \mathbf{R}(\bx) - gbz - p(1-D^{-1})\right) d^3x
+  \int\Scp{\pi}{\p_t b + \bU\cdot \nabla b }_V dt
\,,\label{EB-Lag}
\end{align}
with rotation vector potential $\mathbf{R}(\bx)$ satisfying ${\rm curl}\mathbf{R}(\bx) = 2 {\bm \Omega}(\bx)$. This formula provides the variational derivatives which go into the motion equations in \eqref{EB-PVcons-calc1}. 

For this case, the general equations in \eqref{gen-eqns} become, e.g., for the Euler--Boussinesq equations, 
with $a=(b,D)$, we choose to modify only the advected buoyancy equation, as in \cite{GM1990,GM1996}. Consequently, one finds
\begin{align}
\begin{split}
\p_t  \frac{\de \ell}{\de u} + \L_U  \frac{\de \ell}{\de u}  
&=
D d \frac{\de \ell}{\de D} - \frac{\de \ell}{\de b}db 
+ \left(\frac{\de u^*}{\de b} \cdot  \frac{\de \ell}{\de u} \right) db
\,,\\
\p_t b + \bU\cdot \nabla b &= 0
\quad\hbox{and}\quad 
\p_t D + {\rm div}(D\bU) = 0
\,,
\\& \hbox{with}\quad 
\bU = \bu + \bu^*(b)
\,.\end{split}
\label{EB-example}
\end{align}
Thus, the particle momentum density, mass density and buoyancy are all transported by the sum $\bU = \bu + \bu^*(b)$ of the flow velocity and the bolus velocity. 

Note that the quantity $\frac{\de \ell}{\de u}=\frac{\de \ell}{\de \bu}\cdot d\bx \otimes d^3x$ is a 1-form density, while $\gamma\in V^*$ introduced in equation \eqref{tally-var} lies in the dual space of the advected quantity $a\in V$. The  pressure $p$ in equation \eqref{EB-Lag} is a Lagrange multiplier which enforces $D-1=0$. This constraint leads via the continuity equation to incompressibility of the \emph{augmented} velocity $\bU = \bu + \bu^*(b)$. Because the GM choice for $\bu^*(b)$ in equation \eqref{GM-BV} is \emph{already} divergence-free, the pressure $p$ can then be determined from the motion equation by preservation of the divergence-free condition ${\rm div}\bu=0$, in the usual way, for appropriate boundary conditions. The divergence-free condition for a bolus velocity $u^*(a)$ depending on any other advected quantities besides the buoyancy would also be required for the theory to close. 


Useful formulas for putting the general equations \eqref{gen-eqns} into familiar calculus form for this example are,
\begin{align}
\begin{split}
\L_U (\bv\cdot d\bx) &= \Big( - \bU \times {\rm curl}\bv + \nabla (\bU \cdot \bv) \Big)\cdot d\bx
=  \Big( (\bU \cdot \nabla) \bv + v_j\nabla U^j \Big)\cdot d\bx
\,,\\
\L_U (D\,d^3x) &= {\rm div}(D\bU)\,d^3x
\,,\quad
\L_U b = \bU\cdot\nabla b
\quad\hbox{and}\quad
\bv = \bu + \bR(\bx)
\,.
\end{split}
\label{FIFD}
\end{align}
These formulas allow one to write the VGM EB motion equation in \eqref{EB-example} in standard hydrodynamics form as
\begin{align}
\begin{split}
\p_t \bu + (\bu\cdot \nabla)\bu  &- \bu\times 2{\bm \Omega} + \nabla p + gb\nabla z 
\\ &=
\bu^*(b) \times {\rm curl}\bv - \nabla (\bu^*(b) \cdot \bv) 
+
\left(\frac{\de \bu^* (b)}{\de b}\cdot {\bf v}\right)\nabla b
\,,
\end{split}
\label{VGM-EBmot}
\end{align}
with GM bolus velocity in \eqref{GM-BV}. On the right-hand side of \eqref{VGM-EBmot} three additional
forces appear, all of which are bi-linear in the bolus velocity and the total circulation velocity, $\bv$. First, a Lorentz-type force appears, which is reminiscent of the Craik--Leibovich `vortex force' in the study of Langmuir circulations. Here, the bolus velocity plays the role of the particle velocity in the Lorentz force.  Second, a kinetic pressure force appears depending on higher order gradients of the buoyancy. Third, the action of the differential operator in \eqref{Bolus-op} on the total circulation velocity, $\bv={\bf m}/D$ contributes a force along the buoyancy gradient. 

The first two terms on the right-hand side of equation \eqref{VGM-EBmot} can be combined as we did in equation \eqref{delta-US-V1} of Remark \ref{CL-remark} to compare the Stokes drift $\mathbf{\ob{u}}_S$ in the Craik--Leibovich (CL) theory with the pseudovelocity $\ob{\bf p}/\Dt $ in GLM. Namely,  we compare 
\begin{align}
\big(\, \overline{\xi^j\partial_j \mathbf{u}^\ell} + \overline{{u}_j^\ell \nabla \xi^j} \big)\cdot d\mathbf{x}
= \overline{\mathcal{L}_{\xi}\big(\mathbf{u}^\ell \cdot d\mathbf{x}\big)}
\quad\Longleftrightarrow\quad
\big((\bu^*(b) \cdot \nabla) \bv + v_j\nabla u^{*\,j}(b)\big)\cdot d\bx
=
\mathcal{L}_{u^*(b)}\big(\bv \cdot d\mathbf{x}\big)
\,.\label{bolus-diff-remark}
\end{align}
This relation affords a comparison among the Kelvin circulation theorems for the CL, GLM and VGM theories. In the CL and GLM theories, the Lagrangian mean velocity transports the corresponding Kelvin circulation integrands which contain additional contributions from the fluctuations. However, in the VGM theory the flow circulation is transported by the sum $\bU = \bu + \bu^*(b)$ of the flow velocity and the bolus velocity. 

Thus, the GM model contribution is in the Kelvin loop velocity, while the model contributions in CL and GLM are in the corresponding Kelvin circulation integrands. 

Next, we survey the solution properties of the class of EB VGM equations.

%
%
%

\subsubsection{Kelvin circulation theorem}

The Kelvin circulation theorem for these equations is 
\begin{align}
\frac{d}{dt}\oint_{c(\bU)} \frac{1}{D} \frac{\de \ell}{\de u}
=
- \oint_{c(\bU)} \frac{1}{D} \frac{\de \ell}{\de b}db
+
\oint_{c(\bU)} \frac{1}{D}  \left(\frac{\de u^*}{\de b} \cdot  \frac{\de \ell}{\de u} \right) db
\,.\label{EB-KelThm}
\end{align}
Here the circulation loop moves with the sum of the fluid velocity and the \emph{bolus velocity}, $\bU = \bu + \bu^*(b)$. 

\textbf{Proof.}
Relation \eqref{EB-KelThm} appears, upon substituting the right-hand side of the motion equation in \eqref{EB-example} into the following relation
\begin{align}
\frac{d}{dt}\oint_{c(\bU)} \frac{1}{D} \frac{\de \ell}{\de u}
=
\oint_{c(\bU)}(\p_t + \L_U) \frac{1}{D} \frac{\de \ell}{\de u}
\,.\label{EB-KelThm-proof}
\end{align}
The integration of the pressure gradient(s) in \eqref{EB-example} around the circulation loop vanishes, and the remainder recovers equation  \eqref{EB-KelThm}. 

\subsubsection{PV conservation} 
Potential vorticity (PV) is conserved, since
\begin{align}
\p_t q + \bU\cdot \nabla q = 0
\,\quad\hbox{with}\quad 
q := D^{-1} \nabla b \cdot {\rm curl}\bv
\,\quad\hbox{and}\quad 
\bv = \frac{1}{D} \frac{\de \ell}{\de \bu}
\,.
\label{EB-PVcons}
\end{align}
That is, the PV is conserved along characteristic curves (Lagrangian advection paths) of the sum of the fluid velocity and the bolus velocity.

{\bf Proof.}
The proof can be accomplished either by using the Stokes theorem in the Kelvin theorem \eqref{EB-KelThm}, or perhaps more explicitly, by first casting the EB-type equations in \eqref{EB-example} into a convenient form for taking differentials, as 
\begin{align}
(\p_t + \L_U) (\bv\cdot d\bx) = d \frac{\de \ell}{\de D} - \frac{1}{D} \left( \frac{\de \ell}{\de b} - \gamma \right)db
\,\quad\hbox{and}\quad  
(\p_t + \L_U) db = 0
\,,
\label{EB-PVcons-calc1}
\end{align}
where we have used commutation of differential $d$ and Lie derivative $\L_U$ in taking the differential of the $b$-equation. 

Now taking the differential of the $(\bv\cdot d\bx)$-equation and using $d(\bv\cdot d\bx)= {\rm curl}\bv\cdot d\mathbf{S}$ yields 
\begin{align}
(\p_t + \L_U) \Big(d(\bv\cdot d\bx) \wedge db\Big) = 0
\,.
\label{EB-PVcons-calc2}
\end{align}
Then, using the $D$-equation as $(\p_t + \L_U)D=0$ yields the PV conservation equation in \eqref{EB-PVcons}. 

\subsubsection{Energetics in the Hamiltonian formulation} 

The Legendre transform of the constrained Lagrangian produces an extra term in the Hamiltonian 
\[
h_{GM}(\bfm,a) = h(\bfm,a) + \int \bfm\cdot \bu^*(a) \,d^3x
\,\quad\hbox{with}\quad  
\bfm := \frac{\de \ell}{\de \bu}= D \big(\bu(\bx,t) + \mathbf{R}(\bx)\big)  = D \bv
\,,
\]
with $a=(D,b)$ for the Euler--Boussinesq Hamiltonian
\[
h(\bfm,a) = h(\bfm,D,b) = \int  \frac{1}{2D}|\bfm - \mathbf{R}(\bx)|^2 + Dgbz + p(D-1)\,d^3x
\,.
\]
The semidirect-product Lie--Poisson bracket for the Euler--Boussinesq equations remains the same.  Hence, 
the following Hamiltonian formulation of the GM transport scheme results, for the choice that the bolus velocity depends only on the advected buoyancy variable $b$ and its derivatives, as follows
\begin{align}
\frac{\p }{\p t}\!
\begin{bmatrix}\,
m_j\\ D \\ \ b
\end{bmatrix}
= - 
   \begin{bmatrix}
  m_k\partial_j + \partial_k m_j &
   D\partial_j &
   - \,b_{,j}  &
   \\
   \partial_k D & 0 & 0 & 
   \\
   b_{,k} & 0 & 0 &  
   \end{bmatrix}
   \begin{bmatrix}
{\delta h_{GM}/\delta m_k} = u^k  + u^{*k}(b) \\
{\delta h_{GM}/\delta D} =  \frac12|\bu|^2 + p + gzb  \\
{\delta h_{GM}/\delta b} = D\,gz  -  \left(\frac{\de u^*}{\de b} \cdot  m\right) \\
\end{bmatrix}
\,.
  \label{GM-SD-LPbrkt-det}
\end{align}
The Poisson bracket for this Hamiltonian formulation of the GM transport scheme may be expressed as 
\begin{align}
\frac{d f}{dt}
=
\big\{ f\,,\, h_{GM}\big\}(m,D,b) 
 = - \int
   \begin{bmatrix}
{\delta f/\delta m_k}  \\
{\delta f/\delta D}  \\
{\delta f/\delta b}  \\
   \end{bmatrix}^T
   \begin{bmatrix}
  m_k\partial_j + \partial_k m_j &
   D\partial_j &
   - \,b_{,j}  &
   \\
   \partial_k D & 0 & 0 & 
   \\
   b_{,k} & 0 & 0 &  
   \end{bmatrix}
   \begin{bmatrix}
{\delta h_{GM}/\delta m_k}  \\
{\delta h_{GM}/\delta D}  \\
{\delta h_{GM}/\delta b}   \\
\end{bmatrix}
d^3x
\,.
  \label{GM-SD-LPbrkts}
\end{align}
For $f=h_{GM}$, we find energy conservation, $dh_{GM}/dt=0$, by antisymmetry of the Lie--Poisson bracket in \eqref{GM-SD-LPbrkts}. 

\subsection{Stochastic Closure 2: variational formulation of GM transport}\label{GM-SALT-subsec}

\subsubsection{Stochastic VGM equations}

This section makes a stochastic modification of the variational Gent--McWilliams equations \eqref{gen-eqns}, by taking  the bolus velocity to be stochastic in the Stratonovich sense. Namely, 
\begin{align}
\begin{split}
{\rm d} \frac{\de \ell}{\de u} + \L_U  \frac{\de \ell}{\de u}  
&=
\frac{\de \ell}{\de a}\diamond a\, dt - \left(\frac{\de u^*}{\de a} \cdot  \frac{\de \ell}{\de u} \right) \diamond a \circ dW_t
\,,\\
{\rm d} a + \L_U a &= 0
\,\quad\hbox{with}\quad 
U \to {\rm d}x_t := u(t,x_t)\,dt +u^*(a(x_t))\circ dW_t
\,.\end{split}
\label{gen-eqns-stoch}
\end{align}
Here, the differential notation ${\rm d}x_t$ refers to stochastic evolution of the Lagrangian 
trajectory $x_t=\phi_t(x_0)$ with $\phi_{t=0}=Id$. This stochastic version of the VGM transport scheme also appears in Model 3 of \cite{FGB-Ho2018}, although that paper did not explicitly allow the bolus velocity to depend on gradients of the advected quantities. The difference between the present scheme and the strategy of simply taking the bolus velocity to be stochastic appears in the stochastic term of the motion equation in \eqref{gen-eqns-stoch}. Otherwise the modelling assumptions agree with those in \cite{GM1996}, although they are implemented stochastically and variationally.


\subsubsection{Stochastic Hamiltonian formulation for the GM transport scheme} 

The Legendre transform of the constrained Lagrangian produces an extra term in the Hamiltonian 
\begin{align}
h(\bfm,a) \to {\rm d}h(\bfm,a) =  h(\bfm,a)\,dt + \int \bfm\cdot \bu^*(a) \,d^3x \circ dW_t
\label{gen-Ham-stoch}
\end{align}
with
\[
\bfm := \frac{\de \ell}{\de \bu}= D \big(\bu(\bx,t) + \mathbf{R}(\bx)\big)  =: D \bv
\,.
\]
The semidirect-product Lie--Poisson bracket remains the same.  However, now the 
transport velocity vector field is stochastic,
\[
\frac{\de \,{\rm d}h}{\de \bfm} = {\rm d}x_t := u\,dt +u^*(a)\circ dW_t
\,.
\]
Consequently, we find the following stochastic VGM transport equations for the Euler--Boussinesq equations, when the advected quantity is chosen to be the buoyancy, $b$, as for \cite{GM1990,GM1996},
\begin{align}
{\rm d}\!
\begin{bmatrix}\,
m_j\\ D \\ \ b
\end{bmatrix}
= - 
   \begin{bmatrix}
  m_k\partial_j + \partial_k m_j &
   D\partial_j &
   - \,b_{,j}  &
   \\
   \partial_k D & 0 & 0 & 
   \\
   b_{,k} & 0 & 0 &  
   \end{bmatrix}
   \begin{bmatrix}
{\delta \,{\rm d}h/\delta m_k} ={\rm d}x^k_t := u^k dt  +  u^{*k}(b) \circ dW_t \\
{\delta \,{\rm d}h/\delta D} =  \Big(  \frac12|\bu|^2 + p + gzb \Big)dt  \\
{\delta \,{\rm d}h/\delta b} = D\,gz  \,dt  -  \left(\frac{\de u^*}{\de b} \cdot  m\right) \circ dW_t\\
\end{bmatrix}
.
  \label{Fluid-SD-LPbrkt-stoch}
\end{align}
As in the deterministic VGM transport scheme in the previous section, the constraint $D-1=0$ enforces ${\rm div}(\bu\, dt  +  \bu^*(b) \circ dW_t)=0$ in the stochastic case, as well. This implies ${\rm div}\bu=0$, since we already have ${\rm div}\bu^*(b)=0$ by \eqref{GM-BV}. This result makes the determination of the pressure $p$ systematic and straightforward for stochastic VGM, as well. 

\begin{remark}[Noether's theorem]\rm
The presence of explicit time and space dependence in the stochastic part of the Hamiltonian ${\rm d}h(\bfm,a) $ in \eqref{gen-Ham-stoch} precludes conservation of energy and momentum in the VGM transport scheme, respectively. However, the Kelvin circulation theorem in equation \eqref{EB-KelThm} and the PV conservation in equation \eqref{EB-PVcons} both still persist in the presence of the stochastic transport, modulo replacement of the deterministic advective transport velocity by its stochastic counterpart. These two conservation laws result from Noether's theorem for invariance under relabelling of Lagrangian particles and conservation of advected quantities along Lagrangian particle trajectories. To the extent that the initial spatial distributions of the advected quantities reduce the relabelling symmetry to the isotropy subgroup of the diffeomorphisms which preserves the initial distributions of advected quantities, the Kelvin circulation integral is not preserved in time. The Kelvin--Noether theorem for the Euler--Poincar\'e equation developed in \cite{HMR1998} represents the evolution of the Kelvin circulation resulting from breaking the relabelling symmetry. This is the converse of the Noether theorem for fluid dynamics with advected quantities.

The Legendre transform of the constrained Lagrangian \eqref{EB-Lag} in Hamilton's principle for the Euler--Boussinesq equations, for example, produces the extra term in the Hamiltonian in \eqref{gen-Ham-stoch}. Thus, the additional transport velocity introduced in the advective constraint on the variations in Hamilton's principle \eqref{EB-Lag} is responsible for the choice of the Hamiltonian in equation \eqref{gen-Ham-stoch}. This extra transport velocity is also responsible for the additional forcing of the circulation in equation \eqref{EB-KelThm}.
\end{remark}



\section{Conclusion}\label{conclude-sec}

Motivated by the challenge to create consistent theories of mesoscale and sub-mesoscale wave--current interaction (WCI) discussed in the Introduction, the investigation here began by reviewing GLM, as guided by its WKB formulation in \cite{GjHo1996} for wave packets, in which GLM may be closed at various asymptotic orders. These basic results were reviewed from the viewpoint of geometric mechanics, particularly via the Euler--Poincar\'e formulation of Lagrangian reduction by the symmetry of particle relabelling for continuum mechanics in \cite{HMR1998}. In the geometric mechanics framework, the Lie--Poisson structure of GLM emerges as a classical Hamiltonian field theory with particle  relabelling symmetry. However, the theory is not closed until further assumptions have been made about the group velocity of the waves and the solution for the pressure due to fluctuations. 

Several closure procedures have been introduced previously. In the WKB representation of WCI interaction in Euler--Boussinesq fluids \cite{GjHo1996}, the closure was supplied at various asymptotic orders via the dispersion relation and phase-averaged pressure contributions of the waves. 
By applying slow manifold reduction \cite{MacKay2004} to dynamics in the space of loops, a broader class of variational nonlinear WKB closures for WCI in plasmas was derived in \cite{BurbyRuiz2019}, and expressed in the standard Eulerian frame, rather than the displaced GLM Eulerian frame.
In previous work, similar ideas were applied in both turbulence modelling \cite{HT2012a,HT2012b} and in shape analysis \cite{Bruveris-etal2011}. 
In earlier work on fluid turbulence modelling, a similar type of closure was based on invoking the Taylor hypothesis, that fluctuating quantities would be carried along in the fluid, e.g., \cite{Holm2002a,Holm2002}. 

In the geometric mechanics setting here, we have added considerations of stochastic modelling of the indeterminate quantities in GLM, based on recent advances in stochastic transport \cite{BdLHoLuTa2019}, stochastic variational principles and the Hamiltonian formulations of their results \cite{Holm2015,FGB-Ho2018,HolmRT2018}. 
This variational stochastic formulation seems to promise many future opportunities for the combination of stochastic variational modelling and data assimilation, which in this setting has already had promising results, both in mathematical analysis \cite{CrFlHo2018} and in uncertainty quantification \cite{CoCrHoPaSh2018a, CoCrHoPaSh2018b}. In particular, the analysis in \cite{CrFlHo2018} showed that the presence of the stochastic transport in Euler's fluid equation preserves its analytical properties in the deterministic case. Namely, the stochastic transport version of Euler's fluid equation has local-in-time existence and uniqueness, while also satisfying the Beale--Kato--Majda criterion for blow-up of the solution. 


Section \ref{SPDEs-sec} considers  data-driven and model-driven classes of stochastic closure options for GLM. The purpose of these stochastic closures would be to provide an estimate of the uncertainty in the model computations, which in turn would provide opportunities for reduction of uncertainty by using data assimilation. The data-driven closure option invokes the SALT method of  \cite{CoCrHoPaSh2018a, CoCrHoPaSh2018b}, while the model-driven closure option invokes the familiar Gent--McWilliams approach, as generalized to the stochastic case in \cite{FGB-Ho2018}. 

Because of the close relation of wave propagation to buoyancy dynamics, we chose the stochastic Gent--McWilliams approach in Section \ref{SPDEs-sec} to illustrate the example of stochastic transport in the Euler--Boussinesq equations, rather than taking the full GLM equations. One may regard the GM discussion as a first step toward making the buoyancy dynamics in the wave components of GLM fully stochastic, in the sense of making the noise--mean flow interaction dynamical. 

The GM step also opens the opportunity to quantify the uncertainty of the GM transport scheme, itself. The GM scheme is widely-used in computational ocean science \cite{Gent2011}. Here we note that applying either the deterministic or stochastic GM advective transport scheme in the buoyancy equation in computational simulations while neglecting both its contributions in the motion equation and in the modified incompressibility condition imposed via the continuity equation could be expected to produce errors in the momentum balance. In turn, these errors will cascade into errors in the circulation and PV transport.  It would be interesting to quantify the effects of those types of uncertainties, as well. 

Finally, the introduction of stochastic channels into WCI may provide a means of parameterizing wave breaking. For example, in the GLM setting, under wave forcing at the surface, one could introduce a jump process which would stochastically transfer a certain amount of pseudomomentum $\pb$ to material momentum  $\ob{\bf m}$ while keeping the sum of the two momenta $\pb+\ob{\bf m}$ constant at a given point. For example, this sort of bursting event in momentum transfer could be triggered by a threshold in \emph{wavenumber steepness} $(\bm{\widehat{z}}\cdot \nabla k)^2/|\bm{\widehat{z}}\times\nabla k|^2>1$,  where $k=|{\bf k}| = |\pb|/N$. GLM wave breaking has not been widely considered, and this approach to it has not been tried in applications yet. 
Likewise, in the stochastic GM setting, since the bolus velocity $u^*(b)$ figures dynamically in both the buoyancy equation and the momentum equation in \eqref{Fluid-SD-LPbrkt-stoch}, one might consider jump processes which induce stochastic impulses into the momentum balance which are triggered by the steepness of the local isopycnal slope ${\bf s}=-(\nabla_H b)/ b_z$. Thus, the loss of momentum conservation due to the spatial dependence of the noise would be regarded as stochastic forcing.
These are only preliminary thoughts which must continue to develop and be investigated elsewhere. 

\subsection*{Acknowledgements}
We are grateful for useful referee comments and thoughtful discussions and correspondence about stochastic closures for GLM and wave--current interaction with J. W. Burby, C. J. Cotter, D. Crisan, F. Gay-Balmaz, M. Ghil, J. D. Gibbon, P. Korn, V. Lucarini, E. Luesink, J. C. McWilliams, C. Tronci and B. A. Wingate. This work was partially supported by EPSRC Standard grant EP/N023781/1. 


\appendix

\section[Survey of results for the GLM Euler--Boussinesq stratified fluid]{Survey of results for the GLM Euler--Boussinesq stratified fluid}\label{EBfluid-sec}

The GLM decomposition of the standard Lagrangian in Hamilton's principle 
for an Euler--Boussinesq stratified fluid is given by
{
\begin{align}
\begin{split}
\ell
(\mathbf{u}^\xi, D^\xi, b^\xi,\xi,\partial_t \xi)
&=
\int\bigg\{
D^\xi
\bigg[ \frac{1}{2} \big| \bu^\xi \big|^2
\,
+
\mathbf{R}^\xi \cdot \bu^\xi
-\
\Phi(\mathbf{x}^\xi)
-\,
g\,z\,b^\xi
\bigg]
-
p^\xi
\bigg(
D^\xi
-
1
\bigg)
\bigg\}
d^3x^\xi
\,,
\end{split}
\label{GLM-Bouss-Lag}
\end{align}
}
where $\Phi(\mathbf{x}^\xi)$ is a potential for external or centrifugal forces.
If desired,
the rotation frequency can be allowed to depend on position
along the fluctuating path $\mathbf{x}^\xi$ as
$2\Omega(\mathbf{x}^\xi) = ({\rm curl}\,\mathbf{R})^\xi$. The corresponding 
rotation potential is decomposed in standard GLM fashion as 
$\mathbf{R}^\xi = \mathbf{R}(\mathbf{x}^\xi)=\overline{\mathbf{R}}^L(\mathbf{x}) + \mathbf{R}^\ell(\mathbf{x})$. 

Upon substituting the defining relation 
\begin{align}
\mathbf{u}^\xi := \overline{\mathbf{u}}^L + \frac{D^L\xi}{Dt} = \overline{\mathbf{u}}^L +\mathbf{u}^\ell \,,
\label{u-ell-def}
\end{align}
into \eqref{u-ell-def},  the definition of $\widetilde{D}$ in \eqref{Dtilde-def} allows one to write the corresponding Eulerian mean expression of the averaged Lagrangian for the Euler--Boussinesq stratified fluid as 
{
\begin{align}
\begin{split}
\overline{\ell}
(\overline{\mathbf{u}}^L,\widetilde{D},\overline{b}^L,\xi,\partial_t \xi)
&=
\int\bigg\{
\widetilde{D}
\bigg[
\frac{1}{2}
\overline{
\big| \overline{\mathbf{u}}^L + \mathbf{u}^\ell \big|^2}
\,
+
\overline{
(\overline{\mathbf{R}}^L+\mathbf{R}^\ell)
\cdot\Big(\overline{\mathbf{u}}^L + \mathbf{u}^\ell \Big)}
-\
\overline{\Phi(\mathbf{x}^\xi)}
-\,
g\,z\,\overline{b}^L
\bigg]
\\
&\hspace{1cm}
-
\overline{p^\xi
\Big(
\widetilde{D}
-
{\cal J}
\Big)}
+ 
\overline{
\Big( {\bm \varpi} \cdot ( \p_t  {\bm \xi} + (\mathbf{u}^L\cdot \nabla) {\bm \xi} - {\bm u}^\ell)  \Big)
}
\bigg\}
d^3x
\\
&=
\int\bigg\{
\widetilde{D}
\bigg[
\frac{1}{2} |\overline{\mathbf{u}}^L|^2
+
\overline{\mathbf{u}}^L\cdot\overline{\mathbf{R}}^L
+
\frac{1}{2} \overline{|\mathbf{u}^\ell |^2} 
+
\overline{\mathbf{u}^\ell \cdot \mathbf{R}^\ell }
-
\overline{\Phi}^L(\mathbf{x})
-\,
g\,z\,\overline{b}^L
-\
\overline{p}^L
\bigg]
\\
&\hspace{1cm}
+
\overline{
\Big(p^\xi {\cal J}
\Big)}
+ 
\overline{
\Big( {\bm \varpi} \cdot ( \p_t  {\bm \xi} + (\mathbf{u}^L\cdot \nabla) {\bm \xi} - {\bm u}^\ell)  \Big)
}
\bigg\}
d^3x
\,.
\end{split}
\label{GLM-Bouss-Lag-bar}
\end{align}
}
Here, the last term introduces the Lagrange multiplier ${\bm \varpi}$ to impose 
the constraint that the fluctuation velocity ${\bm u}^\ell$ must satisfy its definition
via the material derivative of the fluctuation vector displacement field ${\bm \xi}$ 
in equation \eqref{u-ell-def}.

The relative buoyancy defined by the mass density ratio $b^\xi=(\rho_{ref}-\rho^\xi)/\rho_{ref}$ is advected as a scalar in the Boussinesq approximation, 
\[
\partial_t\,{b}^\xi+\mathbf{u}^\xi\cdot\nabla{b}^\xi=0
\,,
\]
so we have already substituted
${b}^\xi=\overline{b}^L$ into the Lagrangian in \eqref{GLM-Bouss-Lag-bar}. 
Finally, the
pressure $p^\xi$ in \eqref{GLM-Bouss-Lag} is a Lagrange multiplier that imposes volume preservation
inherited from \eqref{GLM-Bouss-Lag}  via the transformations leading to the
Eulerian average of the constraint relation \eqref{DensityConstraint} defining the conserved
GLM density $\widetilde{D}d^3x=\overline{D^\xi\,d^3x^\xi}= \overline{D^\xi{\cal J}}d^3x$, in the case that $D^\xi=1$.


Most of the important properties of the GLM equations
are discussed in Andrews \& McIntyre [1978a,1978b] \cite{AM1978a,AM1978b}. Many of these
properties arise from general mathematical structures that are shared
by all exact nonlinear ideal fluid theories; namely, as an {\bf Euler-Poincar\'e (EP)
equation} \cite{HMR1998},
{
\begin{equation}\label{EP-GLM-eqn}
\frac{\partial}{\partial t}
\frac{\delta \overline{\ell}}{\delta \overline{u}^L_i}
+\,
\frac{\partial}{\partial x_k}
\Big(\frac{\delta \overline{\ell}}{\delta \overline{u}^L_i}\overline{u}^L_k\Big)
+\,
\frac{\delta \overline{\ell}}{\delta \overline{u}^L_k}
\frac{\partial \overline{u}^L_k}{\partial x_i}
=
\widetilde{D}\frac{\partial}{\partial x_i}
\frac{\delta \overline{\ell}}{\delta \widetilde{D}}
-
\frac{\delta \overline{\ell}}{\delta \overline{b}^L}
\frac{\partial \overline{b}^L}{\partial x_i}
\,,
\end{equation}
}
which is expressed in terms of variational derivatives of an averaged
Lagrangian,  $\overline{\ell}(\overline{\mathbf{u}}^L,\widetilde{D},\overline{b}^L)$ 
and obtained from Hamilton's principle for the Lagrangian mean variables,
\[
0 = \delta S = \delta \int_0^T \overline{\ell}(\overline{\mathbf{u}}^L,\widetilde{D},\overline{b}^L)\,dt
\,.\]
See
Holm, Marsden \& Ratiu  \cite{HMR1998} for an exposition of the
mathematical structures which arise in the EP theory of
ideal fluids which possess advected quantities, such as buoyancy, entropy and magnetic field.
In equation \eqref{EP-GLM-eqn}, for example, the right-hand-side is the usual baroclinic source term.

In particular, the EP equation \eqref{EP-GLM-eqn} for GLM implies the following
Kelvin circulation theorem for the GLM Euler--Boussinesq flow, 
\begin{equation}
\frac{d}{dt} \oint_{\ob\gamma^L(t)} 
\frac{1}{\wt{D}}
\frac{\delta \overline{\ell}}{\delta \ob{\bf u}^L} \cdot  d\mathbf{x} 
=
\oint_{\ob\gamma^L(t)} \left(\nabla \frac{\delta \overline{\ell}}{\delta \widetilde{D}} \cdot d\bx
- \frac{1}{\wt{D}}\frac{\delta \overline{\ell}}{\delta \ob{b}^L} d \ob{b}^L \right),
\label{EP-GLM-KelThm}
\end{equation}
for any closed loop $\ob\gamma^L(t)$ moving with the Lagrangian mean flow velocity $\ob{\bf u}^L$.

The proof of \eqref{EP-GLM-KelThm} follows immediately by noting that 
\begin{equation}
\frac{d}{dt} \oint_{\ob\gamma^L(t)} 
\frac{1}{\wt{D}}
\frac{\delta \overline{\ell}}{\delta \ob{\bf u}^L} \cdot  d\mathbf{x} 
=
\oint_{\ob\gamma^L(t)} 
\Big(\p_t  + \mathcal{L}_{\ob{\bf u}^L}\Big) \left(
\frac{1}{\wt{D}}
\frac{\delta \overline{\ell}}{\delta \ob{\bf u}^L} \cdot  d\mathbf{x} 
\right)
\label{EP-GLM-Lie-form}
\end{equation}
and that the GLM EP equation \eqref{EP-GLM-eqn} may be written as
\begin{equation}
\Big(\p_t  + \mathcal{L}_{\ob{\bf u}^L}\Big) \left(
\frac{1}{\wt{D}}
\frac{\delta \overline{\ell}}{\delta \ob{\bf u}^L} \cdot  d\mathbf{x} 
\right)
=
\nabla \frac{\delta \overline{\ell}}{\delta \widetilde{D}} \cdot d\bx
- \frac{1}{\wt{D}}\frac{\delta \overline{\ell}}{\delta \ob{b}^L} d \ob{b}^L
\,,
\label{EP-GLM-Lie-form}
\end{equation}
after using the advection law for $\wt{D}$ in equation \eqref{contin-eqn}.


\subsubsection*{Variational derivatives and the EP equation for GLM
Euler--Boussinesq stratified fluid}

The mean Lagrangian \[\overline{\ell}\equiv \int\overline{\cal  L}
(\overline{\mathbf{u}}^L,\widetilde{D},\overline{b}^L,\xi,\partial_t \xi)d^3x\]
in equation \eqref{GLM-Bouss-Lag-bar} has been derived via a straight
transcription from the standard Lagrangian for Euler--Boussinesq fluids into
the GLM formalism, followed by taking the Eulerian mean. Its variational derivatives
are given by
\begin{align}\label{GLM-Lag-der}
\begin{split}
\delta\overline{\ell}(\overline{\mathbf{u}}^L,\widetilde{D},\overline{b}^L,\xi,\partial_t \xi)
&=
\int
\bigg[
\Big(\widetilde{D}\big( 
\overline{\mathbf{u}}^L + \overline{\mathbf{R}}^L \big)
+ \overline{\big(\varpi_k \nabla \xi^k\big)}\,\Big)
\cdot\delta\overline{\mathbf{u}}^L
-\
\widetilde{D}\,gz\,\delta\overline{b}^L
-\,\Pi^B\,
\delta\widetilde{D}
\\
&\hspace{1cm}
+ \overline{ \Big( \widetilde{D}\,\big(
\mathbf{u}^\ell + \mathbf{R}^\ell \big) - {\bm \varpi}
\Big)\cdot \de\mathbf{u}^\ell }
+
\overline{
\Big( \de {\bm \varpi} \cdot ( \p_t  {\bm \xi} + (\overline{\mathbf{u}}^L\cdot \nabla) {\bm \xi} - {\bm u}^\ell)  \Big)
}
\\
&\hspace{1cm}
-\,
\overline{
\Big(
\Big( \p_t\varpi_k  + {\rm div}(\varpi_k \overline{\mathbf{u}}^L) + \p_j (p^\xi K^j_k ) \Big) \, \de \xi^k
\Big)}
\,\bigg]d^3x
\,.
\end{split}
\end{align}
The last term in the $\varpi_k$ equation arises from a spatial integration by parts of 
the variation $\overline{p^\xi \,\delta\!\!{\cal J}}$ 
in which $\delta\!\!{\cal J} = K^j_k(\partial\,\delta\xi^k/\partial x^j)$
with cofactor 
\[
K^j_k :={\cal J}({\cal J}^{-1})^j_k
\quad\hbox{with}\quad
{\cal J}^k_j := \frac{\partial \big(x^k+\xi^k(\bx,t)\big) }{ \partial\,x^j}
\,,\quad\hbox{whose determinant is ${\cal J}$.}
\]

Thus, the variations in the fluctuating quantities imply the following quasilinear equations with vanishing mean, 
\begin{align}\label{fluct-vars}
\begin{split}
\de\mathbf{u}^\ell &:\quad \widetilde{D}\,\big(
\mathbf{u}^\ell + \mathbf{R}^\ell \big) - {\bm \varpi} = 0\,;
\\
\de {\bm \varpi} &:\quad \p_t  {\bm \xi} + (\overline{\mathbf{u}}^L\cdot \nabla) {\bm \xi} - {\bm u}^\ell = 0\,;
\\
\de \xi^k &:\quad
\p_t\varpi_k  + {\rm div}(\varpi_k \overline{\mathbf{u}}^L) + \p_j (p^\xi K^j_k ) = 0\,.
\end{split}
\end{align}
The variations with respect to $\delta\overline{\mathbf{u}}^L$ and $\de\mathbf{u}^\ell$ each provides a momentum 
map. Combining them yields,
\begin{equation}
\overline{(\varpi_k \nabla \xi^k)} = \widetilde{D}\,\overline{(u^\ell_k+R^\ell_k)\nabla\xi^k}
=: -\, \pb\,,
\label{pmom-def}
\end{equation}
in which the last step defines the \emph{pseudomomentum density}, ${\pb}$. 
The average of a combination of the second and third equation in \eqref{fluct-vars} will provide the 
dynamical equation we need for the pseudomomentum density in order to close the equations. 
We may also refer to the ratio
\begin{equation}
\overline{\mathbf{v}}:={\pb}/\widetilde{D}:=-\,\overline{(u^\ell_j+R^\ell_j)\nabla\xi^j}
\label{pvel-def}
\end{equation}
as the \emph{pseudovelocity}, $\overline{\mathbf{v}}$, see formula \eqref{circ-trans-pseudomomap}. 

The Boussinesq potential $\Pi^B$ arising in \eqref{GLM-Lag-der} under
the variation of $\overline{\ell}$ with respect to $\widetilde{D}$ is defined by
\begin{equation}\label{Pi-B-def}
\Pi^B
=
\pi^B
+
gz\,\overline{b}^L - \frac{1}{2} |\overline{\mathbf{u}}^L|^2
-
\overline{\mathbf{u}}^L\cdot\overline{\mathbf{R}}^L
\,,
\end{equation}
where
\begin{equation}\label{pi-B-def}
\pi^B
=
\overline{p}^L
+
\overline{\Phi}^L(\mathbf{x})
-
\frac{1}{2} \overline{|\mathbf{u}^\ell |^2} 
-
\overline{\mathbf{u}^\ell \cdot \mathbf{R}^\ell }
\,,
\end{equation}
and, finally, $\overline{p}^L=\overline{p^\xi}$ is the Lagrangian mean pressure.

Upon substituting these variational derivatives into
the  Euler-Poincar\'e (EP) equation (\ref{EP-GLM-eqn}), one finds the
following GLM motion equation governing $\overline{\mathbf{u}}^L$ for a stratified Boussinesq fluid in
Cartesian coordinates, 
\begin{equation}\label{GLM-Bouss-eqns}
\Big[
\frac{D^L}{Dt}\big(
\overline{\mathbf{u}}^L
-
\overline{\mathbf{v}}
\big)
+
\big(
\overline{u}^L_k
-
\overline{v}_k
\big)
\nabla
\overline{u}^L_k
\Big]
- 
\overline{\mathbf{u}}^L\times{\rm curl}\,\overline{\mathbf{R}}^L
+
\nabla\pi^B
+
g\overline{b}^L\boldsymbol{\widehat{z}}
=
0
\,.
\end{equation}
One could also write this equation to mimic a `vortex force'  
in Lorentz form ${\bf E} + {\ob{\bu}}^L\times {\bf B}$ as
\begin{equation}\label{GLM-Bouss-eqns2}
\frac{D^L}{Dt}\overline{\mathbf{u}}^L
+
\frac12 \nabla |{\ob{\bu}}^L|^2
- 
\overline{\mathbf{u}}^L\times{\rm curl}\,\overline{\mathbf{R}}^L
+
\nabla\,\pi^B
+
g\overline{b}^L\boldsymbol{\widehat{z}}
=
\Big(\p_t \overline{\mathbf{v}}
+
\nabla ({\ob{\bu}}^L\cdot {\ob{\bv}})\Big)
- 
\ob{\bu}^L \times {\rm curl}\ob{\bv} 
\,.
\end{equation}
For convenience, the equations for the advected quantities $\overline{b}^L$ and $\widetilde{D}$ are recalled from  
above as
\begin{equation}\label{aux-eqns}
\partial_t\,{\overline{b}^L}+\overline{\mathbf{u}}^L\cdot\nabla{\overline{b}^L}=0
\quad\hbox{and}\quad
\partial_t\,\widetilde{D}
+
{\rm div}(\widetilde{D}\overline{\mathbf{u}}^L )
\,.
\end{equation}

\begin{remark}[Comparison of GLM pseudomomentum dynamics with the Craik-Leibovich theory]\label{CL-remark}\rm
Without the `${\bf E}$-field' term on its right side, equation \eqref{GLM-Bouss-eqns2} would seem to 
have the same form as the Craik-Leibovich theory,
except that the Stokes mean drift velocity $\mathbf{\bar{u}}_S$ would have been replaced by the pseudovelocity $\ob{\bv}$. 
Formally, then, the GLM Euler--Boussinesq stratified fluid equations might appear to comprise a dynamical version of the Craik-Leibovich theory. However, the pseudovelocity $\ob{\bv}$ is by no means the same as the Stokes mean drift velocity, $\mathbf{\ob{u}}_S$. In fact, their difference has nonzero circulation.
This is because  the pseudovelocity, $\ob{\bv}=\ob{\bf p}/\Dt$, and the Stokes mean drift velocity, $\mathbf{\ob{u}}_S$, are complementary quantities in the Eulerian mean of $\mathcal{L}_{\xi}(\mathbf{u}^\ell \cdot d\mathbf{x})$, which is the Lie derivative of the fluctuating circulation 1-form $\mathbf{u}^\ell \cdot d\mathbf{x}$ with respect to the fluctuation vector field, ${\bm \xi}$. Namely,
\begin{equation}\label{delta-US-V1}
\big(\mathbf{\ob{u}}_S - \ob{\bf p}/\Dt \big)\cdot d\mathbf{x}
= \big(\, \overline{\xi^j\partial_j \mathbf{u}^\ell} + \overline{{u}_j^\ell \nabla \xi^j} \big)\cdot d\mathbf{x}
= \big(\, - \,\overline{ \boldsymbol{\xi} \times {\rm curl} \mathbf{u}^\ell} 
+ \nabla\overline{( \boldsymbol{\xi}\cdot \mathbf{u}^\ell )}\,\,\big) \cdot d\mathbf{x}
= \overline{\mathcal{L}_{\xi}\big(\mathbf{u}^\ell \cdot d\mathbf{x}\big)}
\,.\hspace{1cm}  \square
\end{equation}
So the two `velocities' meet here in the Lie derivative. They are so different that their difference means something. The Stokes mean drift velocity, $\mathbf{\ob{u}}_S$, is the rate of distortion of the fluctuating velocity covector by the fluctuating disturbance in the Lagrangian path away from its mean, as if the covector were an array of scalars. The pseudovelocity $\ob{\bv}$ is (minus) the corresponding rate of distortion of its covector basis. The place where all this comes together is in the GLM Kelvin's theorem when we bring in the Eulerian mean velocity $\ob{\bf u}^E$ to transform from Lagrangian mean to Eulerian mean quantities in the integrand as 
\begin{equation}\label{delta-US-V2}
\oint_{c(\ob{\bf u}^L)}(\ob{\bf u}^L - \ob{\bv})\cdot d\bx 
= \oint_{c(\ob{\bf u}^L)}(\ob{\bf u}^E + \ob{\bf u}^S - \ob{\bv})\cdot d\bx
= \oint_{c(\ob{\bf u}^L)} \ob{\bf u}^E\cdot d\bx +  \overline{\mathcal{L}_{\xi}\big(\mathbf{u}^\ell \cdot d\mathbf{x}\big)}
\,.
\end{equation}
For further discussion of the geometric and Hamiltonian properties of the Craik--Leibovich theory, see \cite{HolmCL1996}.
\end{remark}

\begin{remark}\rm
We still need an equation for the pseudomomentum density ${\pb}$ in equation 
\eqref{pmom-def} in order to close the GLM
Euler--Boussinesq motion equation in \eqref{GLM-Bouss-eqns}. 
However, before deriving that equation, let us 
make a few remarks about the properties of the (as yet unclosed) GLM equations for the 
Euler--Boussinesq stratified fluid which have been obtained, so far.
\end{remark}


\subsubsection*{Relation to the EP Kelvin circulation theorem for GLM Boussinesq
stratified fluid}

The GLM average of Kelvin's circulation integral is defined as,
\begin{align}
\begin{split}
\overline{I(t) }
&=
\overline{ \oint_{\gamma^\xi(t)} 
\big(\mathbf{u}^\xi+\mathbf{R}(\mathbf{x}^\xi)\big)
\cdot d \mathbf{x}^\xi 
}
=
\overline{\oint_{\gamma^\xi(t)} 
\big(\overline{\mathbf{u}}^L + \overline{\mathbf{R}}^L
+
\mathbf{u}^\ell +\mathbf{R}^\ell\big) 
\cdot
(d\mathbf{x}+d\xi) }
\\
&=
\oint_{ \bar\gamma^L(t) }
\big(\overline{\mathbf{u}}^L + \overline{\mathbf{R}}^L
+ \overline{(u^\ell_k+R^\ell_k)\nabla\xi^k\big)}
\cdot  d\mathbf{x} 
= 
\oint_{\bar\gamma^L(t)} 
(\overline{\mathbf{u}}^L + \overline{\mathbf{R}}^L - \overline{\mathbf{v}}) 
\cdot  d\mathbf{x} 
\,,
\end{split}
\label{kelnt-psm}
\end{align}
where the contour $\bar\gamma^L(t)$ moves with velocity 
$\overline{\mathbf{u}}^L$, since it follows the fluid parcels as the average
is taken. Thus, the Lagrangian mean leaves invariant the {\it form}
of the Kelvin integral, while averaging the {\it velocity} of its
contour. In addition, the pseudovelocity co-vector $\overline{\mathbf{v}}$ defined
in \eqref{pmom-def}  appears in the \emph{integrand} of the GLM averaged 
Kelvin integral $\overline{I(t) }$.

The time derivative of the GLM averaged Kelvin circulation
integral is, cf. formula \eqref{circ-trans-pseudomomap},
\begin{align}
\begin{split}
\frac{d}{dt}\overline{I(t) }
&=
\oint_{c(\overline{\mathbf{u}}^L)}
\big( \partial_t + \mathcal{L}_{\overline{\mathbf{u}}^L} \big)
\Big( \big(\overline{\mathbf{u}}^L + \overline{\mathbf{R}}^L - \overline{\mathbf{v}}
\big)\Big)
\cdot
d\mathbf{x}
\Big)
\\&=
\oint_{\bar\gamma^L(t)} \!\!
\Big[(\partial_t+\overline{\mathbf{u}}^L\cdot\nabla)
(\overline{\mathbf{u}}^L - \overline{\mathbf{v}}) 
+
(\overline{u}^L_k - \overline{v}_k) \nabla \overline{u}^{L\,k}
+
2\Omega\!\times\!\overline{\mathbf{u}}^L
+ \nabla \big(\overline{\mathbf{u}}^L\cdot \overline{\mathbf{R}}^L(\mathbf{x}) \big)
\Big]
\!\!\cdot\!d\mathbf{x} 
\,.
\end{split}
\label{kel-dot}
\end{align}
where ${\rm curl}\,\overline{\mathbf{R}}^L(\mathbf{x})
= 2\Omega(\mathbf{x})$.
The combination of terms in the integrand defines the {\bf transport
structure} of the GLM theory under the Lie derivative $\mathcal{L}_{\overline{\mathbf{u}}^L} $
along the mean velocity vector, $\overline{\mathbf{u}}^L$. From the GLM motion equation
\eqref{GLM-Bouss-eqns} one now finds the GLM Kelvin circulation theorem for 
Boussinesq incompressible flow,
\begin{equation}\label{GLM-comp-Kel-thm}
\frac{d}{dt}\overline{I(t) }
=
\frac{d}{dt}\oint_{c(\overline{\mathbf{u}}^L)}
\Big( \overline{\mathbf{u}}^L + \overline{\mathbf{R}}^L - \overline{\mathbf{v}}
\Big)
\cdot
d\mathbf{x}
= -g\,
\oint_{c(\overline{\mathbf{u}}^L)}
\overline{b}^L \,dz
\,.
\end{equation}
\begin{remark}\rm
Thus, the Lagrangian mean {\it averages the velocity} of the fluid
parcels on the Kelvin circulation loop, while it {\it adds the mean
contribution} of the velocity fluctuations to the integrand of the Kelvin circulation.
\end{remark}

Equation \eqref{EP-GLM-eqn} in the EP framework provides the {\bf Kelvin-Noether
theorem} for Boussinesq stratified fluid, in the form
\begin{equation}
\frac{d}{dt}\oint_{c(\overline{\mathbf{u}}^L)}
\frac{1}{\widetilde{D}}
\frac{\delta \overline{\ell}}{\delta \overline{\mathbf{u}}^L}
\cdot
d\mathbf{x}
=
-\,\oint_{c(\overline{\mathbf{u}}^L)}
\frac{1}{\widetilde{D}}
\frac{\delta \overline{\ell}}{\delta \overline{b}^L}
\,d\overline{b}^L
.
\end{equation}
Evaluating this for the  GLM Boussinesq stratified fluid with $\ob{\ell}$ given in \eqref{GLM-Bouss-Lag-bar} yields,
\begin{equation}
\frac{d}{dt}\oint_{c(\overline{\mathbf{u}}^L)}
\Big( \overline{\mathbf{u}}^L 
+
\overline{\mathbf{R}}^L(\mathbf{x})
- \ob{\bf v}
\Big)
\cdot
d\mathbf{x}
=
\oint_{c(\overline{\mathbf{u}}^L)}
gz\,d\overline{b}^L
,
\label{GLM-EB-Kelvin-thm}
\end{equation}
which agrees with the result of the direct calculation in \eqref{GLM-comp-Kel-thm}. 

If the loop $c(\overline{\mathbf{u}}^L)$ moving with the Lagrangian mean
flow lies entirely on a level surface of $\overline{b}^L$, then the
right hand side vanishes, and one recovers for this case the
``generalized Charney-Drazin theorem'' for transient Boussinesq
internal waves, in analogy to the discussion in Andrews \& McIntyre
\cite{AM1978a} for the adiabatic compressible case.

{\bf Total vorticity.}
Finally, upon defining the \emph{total vorticity} as 
\begin{equation}\label{GLM-vort1}
{\bm \omega}_{tot} := {\rm curl}\Big( \overline{\mathbf{u}}^L + \overline{\mathbf{R}}^L - \overline{\mathbf{v}}
\Big)
\end{equation}
and applying the Stokes theorem to the GLM Kelvin theorem in equation \eqref{GLM-comp-Kel-thm}, one finds
\begin{equation}\label{GLM-vort2}
\frac{d}{dt}\overline{I(t) }
=
\oint_{c(\overline{\mathbf{u}}^L)}
\big( \partial_t + \mathcal{L}_{\overline{\mathbf{u}}^L} \big)
\big({\bm \omega}_{tot}\cdot d{\bm S}  \big)
=
- \oint_{c(\overline{\mathbf{u}}^L)} g \nabla {\ob b}^L\times \mathbf{\widehat{z}} 
\cdot d{\bm S}\,.
\end{equation}
Since this equation holds for any loop, we have
\begin{equation}\label{GLM-vort3}
\p_t {\bm \omega}_{tot}  - {\rm curl} \big({\ob u}^L\times {\bm \omega}_{tot}\big) = - g \nabla {\ob b}^L\times \mathbf{\widehat{z}}\,.
\end{equation}

\begin{remark}\rm
Thus, the EB vorticity equation keeps its form in the GLM theory, while it {\it adds the mean
contribution} of the velocity fluctuations to the total vorticity defined in \eqref{GLM-vort1}. 
\end{remark}

\subsubsection*{Local potential vorticity conservation for GLM
Boussinesq stratified fluid}

Invariance of the Lagrangian under diffeomorphisms
(interpreted physically as Lagrangian particle relabeling) implies
the local conservation law for EP potential vorticity, 
\begin{equation}
\frac{D^L}{Dt}\overline{q}^L
=
0
\,,\quad\hbox{where}\quad
\overline{q}^L
=
\frac{1}{\widetilde{D}}\nabla\overline{b}^L\cdot
{\rm curl}\,
\Big(
\frac{1}{\widetilde{D}}
\frac{\delta \overline{\ell}}{\delta \overline{\mathbf{u}}^L}\Big)
\,.
\nonumber
\end{equation}
For the GLM case, the potential vorticity is given explicitly as
\begin{equation}
\overline{q}^L
=
\frac{1}{\widetilde{D}}\nabla\overline{b}^L\cdot
{\rm curl}\,
\Big( \overline{\mathbf{u}}^L 
- \overline{\mathbf{v}}
+
\overline{\mathbf{R}}^L(\mathbf{x})
\Big)\,.
\nonumber
\end{equation}
The EP framework explains the relation of the potential
vorticity to the Kelvin circulation theorem. However, there remains the question of the 
evolution of the pseudovelocity, $\overline{\mathbf{v}}$.

\subsubsection*{Fluctuation equations}

Hamilton's principle for the Lagrangian mean variables $\{\overline{\mathbf{u}}^L,\widetilde{D},\overline{b}^L\}$ has already been 
calculated in equation \eqref{GLM-Lag-der}.
We now apply Hamilton's principle for the fluctuation variable $\xi^k$ using the original  Lagrangian $\ell(\mathbf{u}^\xi, D^\xi, b^\xi,\xi,\partial_t \xi)$ in equation \eqref{GLM-Bouss-Lag}.
\[
0 = \delta S = \delta \int_0^T \overline{\ell}(\overline{\mathbf{u}}^L,\widetilde{D},\overline{b}^L
,\xi,\partial_t \xi )\,dt
\,.\]
The result for the \emph{momentum density} $\varpi_k$ canonically conjugate to  $\xi^k$ is
\begin{equation}
\varpi_k := \frac{\delta \overline{\ell} }{\delta (\partial_t \xi^k) } 
= \widetilde{D} \Big(\frac{D^L\xi_k}{Dt} + R_k(\mathbf{x}^\xi)\Big)
= \widetilde{D} \Big( u^\ell_k + R_k^\xi \Big)
.
\end{equation}

\paragraph{Wave action density.}
To introduce the wave action density $N$ and explain how it is related to the GLM pseudomomentum density, ${\pb}$, we take the Eulerian mean of the following
pre-canonical transformation,
\[
{\pb}\cdot d\mathbf{x}
=
-\,\overline{\varpi_k \nabla\xi^k}\cdot d\mathbf{x}
=
-\,\overline{\bm{\varpi}\cdot d\bm{\xi} }
\,.\]
If $\xi$ and $\pi$ are averaged over a \emph{phase parameter}, $\phi$, we may write
the phase-averaged differential relation as
\[
{\pb}\cdot d\mathbf{x}
=
-\,\overline{\bm{\varpi}\cdot d\bm{\xi} }
= 
- \overline{\varpi_k\partial_\phi\xi^k} \, d\phi
=
Nd\phi
=
N\mathbf{k}\cdot d\mathbf{x}
\,,\]
where the wavevector $\mathbf{k}$ is defined by $d\phi =
\nabla\phi\cdot d\mathbf{x} = \mathbf{k}\cdot d\mathbf{x}$
and the \emph{wave action density} $N$ is given by 
\[
N =-\,\overline{\varpi_k\partial_\phi\xi^k}
\,.\]

Thus, the wave action density $N =-\overline{\varpi_k\partial_\phi\xi^k}$ is 
related to the GLM pseudomomentum by 
${\pb}=N\mathbf{k}$. 

For the WKB wavepacket 
\[
\xi(\bx,t) = \tfrac{1}{2}(\mathbf{a}(\bx,t)e^{i\phi(\bx,t)/\epsilon}
+
\mathbf{a}^*(\bx,t)e^{-i\phi(\bx,t)/\epsilon})
\,,\] 
one finds the formula for constant Coriolis parameter $2\Omega$, Gjaja \&  Holm \cite{GjHo1996},
\begin{align}
\begin{split}
\frac{N}{\widetilde{D}}
&=
-\,\overline{\Big[
\frac{D^L{\bm\xi}}{Dt}+({\bm \Omega}\times{\bm\xi})\Big]\cdot\partial_\phi{\bm\xi}}
=
-\,\overline{\Big[
(\bu^\ell + \mathbf{R}^\ell)\Big]\cdot\partial_\phi{\bm\xi}}
\\&=\
2\widetilde{\omega}|\mathbf{a}|^2
+
2i{\bm \Omega}\cdot\mathbf{a}\times\mathbf{a}^*
+
2\Im\Big(\mathbf{a}\cdot\frac{D^L\mathbf{a}^*}{Dt}\Big)
\,,
\end{split}
\label{N-def}
\end{align}

in which the quantity
\[
\widetilde{\omega}=-D^L\phi/Dt
=\omega-\mathbf{k}\cdot\overline{\mathbf{u}}^L
\]
is the Doppler-shifted wave frequency.  As a result of the symmetry under
translations in $\phi$ induced by phase-averaging the Lagrangian,
the corresponding Euler--Lagrange equation implies the conservation law,
\begin{equation}\label{Wave-action-cons}
0
=
-\,
\frac{\partial }{\partial t}
\frac{\partial \overline{\cal L}}{\partial (\partial_t\phi)}
-\,
{\rm div}\,
\frac{\partial \overline{\cal L}}{\partial (\nabla\phi)}
=
\frac{\partial }{\partial t}
\frac{\partial \overline{\cal L}}{\partial \omega}
-\,
{\rm div}\,
\frac{\partial \overline{\cal L}}{\partial {\bm k} }
=
\frac{\partial N}{\partial t}
+
\frac{\partial }{\partial x^j}
\Big(N \big(\overline{u}^{L\,j}
+
\overline{\big(p^\xi K^j_i \,\p_\phi\xi^i\big)}\big)\, \Big)
\,,
\end{equation}
upon using the variational derivatives in equation
(\ref{GLM-Lag-der}). Andrews \& McIntyre [1978b] \cite{AM1978b} obtain the same
conservation law by directly manipulating the GLM motion equation. 
This is also Noether's theorem for symmetry of the Lagrangian under phase shifts.
For more discussion from a variational viewpoint in the case that the fluctuations are 
single-frequency wave packets with slowly varying envelopes, see also Gjaja \&  Holm \cite{GjHo1996}. 
Of course, Noether's theorem always applies in averaging Hamilton's principle, since such averaging 
always produces a continuous symmetry of the Lagrangian. In general, Noether's theorem implies the following
about the relation of averaging to local conservation laws,  \cite{Hayes1970, AM1978b, Holm2002a,Holm2002}.
%
\begin{lemma}\rm
When Lagrangian averaging introduces an ignorable coordinate in fluid dynamics, the average of the
corresponding canonically conjugate momentum is locally conserved; 
that is, the corresponding quantity is conserved in a shifted frame of motion relative to Lagrangian fluid parcels.

In this case, the locally conserved quantity is the wave action density $N$ in \eqref{N-def}, which is 
the phase-averaged quantity (momentum map) whose canonical Poisson bracket generates phase shifts.
The spatial integral over the domain of. flow $\int_{\cal D} N\, d^3x$ is conserved globally,
for appropriate boundary conditions.
\end{lemma}

We interpret equation \eqref{Wave-action-cons} as local conservation of wave action $N$, as 
transported by the sum of the mean material velocity and the \emph{relative} group velocity $\overline{v}_G$, defined by
\begin{equation}\label{group-vel-def}
\overline{v}_G^j := \overline{(p^\xi K^j_i \,\p_\phi\xi^i)}
\end{equation}
so that
\begin{equation}\label{N-eqn}
\frac{\partial N}{\partial t}
+
{\rm div}\big(N(\ob{\bu}^L + \ob{\boldsymbol{v}}_G)\big) = 0
\,.
\end{equation}

\paragraph{Pseudomomentum dynamics -- Hamiltonian formulation.}
It remains to determine the dynamical equation for the pseudomomentum $\pb$. For this, we shall 
pass to the Hamiltonian side via the following Legendre transform,
\begin{align}
\begin{split}
\Hbar (\ob{\bf m}, N, \pb, \Dt, \bb^L; \omega, \bk, \overline{\bm v}_G)
&= \int \overline{\mathbf{m}}\cdot \overline{\mathbf{u}}^L 
+ N\omega + ({\pb} - N\bk) \cdot \big( \ob{\bu}^L + \overline{\bm v}_G)   \  d^3x 
- \overline{\ell}(\overline{\mathbf{u}}^L,\widetilde{D},\overline{b}^L,\xi,\partial_t \xi)
\\&= \int \bigg[
\frac{1}{2\widetilde{D}}\big| \mb + \pb - \Dt\Rb^L  \big|^2 
+ \pb \cdot \ob{\bm v}_G  + N\big(\omega - \bk \cdot ( \ob{\bu}^L + \ob{\bm v}_G) \big)
\\&\hspace{1cm}
+ 
\Dt \Big( \pbL + gz\bb^L + \PhibL(\bx) \Big)
- 
\Dt \Big( \frac12 \overline{ | \bu^\ell |^2} +  \overline{ \bu^\ell\cdot\bR^\ell} \Big)
\\&\hspace{1cm}
-
\overline{
\Big(p^\xi {\cal J}
\Big)}
- 
\overline{
\Big( {\bm \varpi} \cdot ( \p_t  {\bm \xi} + (\ob{\mathbf{u}}^L\cdot \nabla) {\bm \xi} - {\bm u}^\ell)  \Big)
}
\,\bigg]\,d^3x
\end{split}
\label{Hbar-det}
\end{align}
We do not vary $\Hbar$ with respect to the parameters $\omega, \bk$ and $\overline{\bm v}_G$. The term $({\pb} - N\bk)\cdot \ob{\bm v}_G $ vanishes for arbitrary $\ob{\bm v}_G$, as a consequence of the variation
in $\ob{\mathbf{u}}^L$. Moreover, the expected `wave conservation relation' $\partial_t \mathbf{k} = -\nabla \omega$ will follow as a result of the other dynamical equations.
We note that the constraints on the averaged Lagrangian $\ob{\ell}$ will still apply for the Hamiltonian, since they are not Legendre transformed. We may now compute the variations of the Hamiltonian as
\begin{align}
\begin{split}
\de\Hbar &= \int \ub^L\cdot \de \mb
+ (\widetilde{D}\,gz)\,\delta\overline{b}^L + \Pi_{tot}\,\de \widetilde{D}
+ \de N\big(\omega - \bk \cdot ( \ob{\bu}^L + \overline{\bm v}_G) \big)
\\&\hspace{1cm}
+ \de \pb\cdot ( \ob{\bu}^L + \overline{\bm v}_G)
 + ({\pb} - N\bk) \cdot \de \ob{\bu}^L  
\,d^3x\,,
\end{split}
\end{align}
where $\Pi_{tot}$ is given by
\begin{align}
\Pi_{tot} = \frac{\de \Hbar}{\de \Dt} 
= \Big(\pbL + gz\bb^L + \PhibL(\bx) \Big)
- \Big( \frac12 \overline{ | \bu^\ell |^2} +  \overline{ \bu^\ell\cdot\bR^\ell} \Big)
=: \overline{\Pi}^L + \overline{\pi^\ell}
\,.
\label{Pi-tot}
\end{align}
Vanishing of the other variations of the averaged Lagrangian $\ob{\ell}$ in \eqref{Hbar-det} still enforces the constraints \eqref{fluct-vars} since the corresponding variables were not Legendre transformed. 

{\bf Wave component.} 
We now write the equations of motion for the pseudomomentum density and wave action density in Lie--Poisson form, following the lead of Gjaja and Holm \cite{GjHo1996}
\begin{align}
\begin{split}
\frac{ \p {\ob{p}_j} }{ \p t} &= \left\{ {\ob{p}_j}\,, \overline{H} \right\}
= -\,(\ob{p}_k\partial_j + \partial_k \ob{p}_j) \frac{\delta \overline{H}}{\delta {\ob{p}_k}}
-\, N \p_j \frac{\delta \overline{H}}{\delta N}
\\&
= - \ (\ob{p}_k\partial_j + \partial_k \ob{p}_j)\big( \overline{u}^{L\,k} + \overline{v}_G^k \big)
- \ N \p_j\big(\omega - \mathbf{k}\cdot (\ob{\bf u}^L+\overline{\boldsymbol{v}}_G)\,\big)\,,
\\
\frac{ \p N }{ \p t} &= \left\{ N\,, \overline{H} \right\}
= - \ \p_k  \left(N \frac{\delta \overline{H}}{\delta {\ob{p}_k}}\right)
= - \ \p_k  \left(N \big( \overline{u}^{L\,k} + \overline{v}_G^k \big)\right)
\,,
\end{split}
\label{mot-eqns}
\end{align}
in which we have used the relations,
\[
\frac{\delta \overline{H}}{\delta {\ob{p}_j}}
 =  \overline{u}^{L\,j} + \overline{v}_G^j 
 \,,\quad
 \frac{\delta \overline{H}}{\delta N} = \omega - k_i (\ob{u}^{L\,i} + \overline{v}_G^i)
\,,\]
and we can may choose $\overline{v}_G^j =  \overline{(p^\xi K^j_i \,{\p_\phi\xi^i}^i)}$ to agree with the definition in \eqref{group-vel-def}. 

\begin{remark}[Wave conservation]\rm $\,$

Note that equations \eqref{mot-eqns} and the relation ${\pb} = N\bk$ imply the wave conservation relation $\partial_t \mathbf{k} = -\nabla \omega$.
\end{remark}

{\bf Lie--Poisson Hamiltonian structure}
The wave field's semidirect-product Lie--Poisson Hamiltonian structure may be revealed by its
Poisson operator, given in matrix form by
\begin{align}
\p_t\!
\begin{bmatrix}\,
\ob{p}_j \\ N 
\end{bmatrix}
= - 
   \begin{bmatrix}
   \ob{p}_k\partial_j + \partial_k \ob{p}_j &  N\partial_j 
   \\
   \partial_k N & 0 &  
   \end{bmatrix}
   \begin{bmatrix}
{\delta \Hbar/\delta \ob{p}_k} = \overline{u}^{L\,k} + \overline{v}_G^k \\
{\delta \Hbar/\delta N} =  \omega - k_i (\ob{u}^{L\,i} + \overline{v}_G^i)
\end{bmatrix}
\,.
  \label{Waves-SD-LPmatrix}
\end{align}
Expanding out the matrix product yields the Lie--Poisson bracket between two functionals $F$ and $H$ as,
\begin{align}
\begin{split}
\frac{d}{dt} F(\ob{\bf p},N)  = 
\Big\{F, H \Big\}&=
- \int 
\begin{bmatrix}
\,
\de F/\de \ob{p}_j \\ \de F/\de N 
\end{bmatrix}^T
   \begin{bmatrix}
   \ob{p}_k\partial_j + \partial_k \ob{p}_j &  N\partial_j 
   \\
   \partial_k N & 0 &  
   \end{bmatrix}
   \begin{bmatrix}
{\delta H/\delta \ob{p}_k} \\
{\delta H/\delta N} 
\end{bmatrix}
d^3x
\\&= - \int 
\frac{\de F}{\de \ob{p}_j} \left((\ob{p}_k\partial_j + \partial_k \ob{p}_j) \frac{\delta H}{\delta \ob{p}_k}
+ 
N\partial_j \frac{\delta H}{\delta N}\right) +  \frac{\delta F}{\delta N}\big(\partial_k N \big)\frac{\delta H}{\delta \ob{p}_k} 
\,d^3x\,.
\end{split}
  \label{Waves-SD-LPbrkt}
\end{align}

The Lie-Poisson bracket in equation \eqref{Waves-SD-LPbrkt} is
defined on the dual of the semidirect-product Lie algebra $\mathfrak{X} \circledS  \Lambda^0$ of vector fields
$X\in\mathfrak{X}(M)$ and functions $f\in\Lambda^0(M)$ on the domain of flow, $M$. The corresponding Lie algebra commutator is given by
\begin{align}
\big[ (X, f) , (\ob{X}, \ob{f}) \big] = \big( [X,\ob{X}]\,,\,X(\ob{f}) - \ob{X}(f)\big)\,,
  \label{SDP-LAbrkt}
\end{align}
where $ [X,\ob{X}]$ is the commutator of vector fields and $X(\ob{f})$ is the Lie derivative of vector fields acting on functions. 
The dual coordinates are: the pseudomomentum 1-form density, $\ob{p}=\pb\cdot d {\bf x}\otimes d^3x$, dual to vector fields;  and the wave action density, $N d^3x$,  dual to functions. Thus, the Lie-Poisson bracket in equation \eqref{Waves-SD-LPbrkt}  may be written as 
\begin{align}
\begin{split}
\Big\{F, H \Big\}(\ob{p},N)&= 
\SCP{( \ob{p},N ) }{\left[ \left(\frac{\de F}{\de \ob{p}}\,,\, \frac{\de F}{\de N}\right)
\,,\,
\left(\frac{\de H}{\de \ob{p}}\,,\, \frac{\de H}{\de N}\right) \right] }_{\mathfrak{X},V}
\\ \\&=
\SCP{\ob{p}}{\left[\frac{\de F}{\de \ob{p}}\,,\, \frac{\de H}{\de \ob{p}} \right]}_\mathfrak{X}
+ \SCP{\L_{\frac{\de F}{\de \ob{p}}} N}{\frac{\de H}{\de N}}_V
- \SCP{\L_{\frac{\de H}{\de \ob{p}}} N}{\frac{\de F}{\de N}}_V
.\end{split}
  \label{Waves-SD-LPbrkt1}
\end{align}
In other standard notation \cite{HMR1998}, this is 
\begin{align}
\begin{split}
\Big\{F, H \Big\}(\ob{p},N)&=
- \SCP{\ob{p}} {{\rm ad}_{ \frac{\de H}{\de \ob{p} }}   \frac{\de F}{\de \ob{p}} }_\mathfrak{X}
+ \SCP{\L_{\frac{\de F}{\de \ob{p}}} N}{\frac{\de H}{\de N}}_V
- \SCP{\L_{\frac{\de H}{\de \ob{p}}} N}{\frac{\de F}{\de N}}_V
\\ \\ &=
- \SCP{ {\rm ad}^*_{ \frac{\de H}{\de \ob{p} }}  \ob{p}} { \frac{\de F}{\de \ob{p}} }_\mathfrak{X}
- \SCP{\frac{\de H}{\de N} \diamond N }{ \frac{\de F}{\de \ob{p}} }_\mathfrak{X}
- \SCP{\L_{\frac{\de H}{\de \ob{p}}} N}{\frac{\de F}{\de N}}_V.
\end{split}
  \label{Waves-SD-LPbrkt2}
\end{align}

The corresponding forms of their equations of motion in \eqref{mot-eqns} when written in terms of Lie derivatives are
\begin{align}
\begin{split}
\big(\p_t + \mathcal{L}_{(\ob{u}^{L} + \ob{v}_G)}\big) \big(\pb\cdot d\bx \otimes d^3x \big)
&=  -\big( N\,d^3x\big) d \big(\omega - k_i (\ob{u}^{L\,i} + \overline{v}_G^i) \big)
\,,\\
\big(\p_t + \mathcal{L}_{(\ob{u}^{L} + \ob{v}_G)}\big) \big( N\,d^3x\big) &= 0
\,.
\end{split}
\label{Waves-LieDerivForm}
\end{align}
Thus, the pseudomomentum density and the wave action density are both transported by the sum of the Lagrangian mean velocity and the group velocity. 

{\bf Material component.} 
The semidirect-product Lie--Poisson bracket for the fluid material component of the flow is also revealed
 by the matrix form of its Poisson operator,
\begin{align}
\p_t\!
\begin{bmatrix}\,
\ob{m}_j\\ \Dt \\ \ \bb^L
\end{bmatrix}
= - 
   \begin{bmatrix}
   \ob{m}_k\partial_j + \partial_k \ob{m}_j &
   \Dt\partial_j &
   - \,\bb^L_{,j}  &
   \\
   \partial_k \Dt & 0 & 0 & 
   \\
   \bb^L_{,k} & 0 & 0 &  
   \end{bmatrix}
   \begin{bmatrix}
{\delta \Hbar/\delta \ob{m}_k} = \ob{u}^{L k} \\
{\delta \Hbar/\delta \Dt} =  \Pi_{tot} \\
{\delta \Hbar/\delta \bb^L} = \widetilde{D}\,gz \\
\end{bmatrix}
.
  \label{Fluid-SD-LPbrkt}
\end{align}
The corresponding Lie--Poisson bracket between two functionals $F$ and $H$ of $\{\ob{\bf m},\Dt,\ob{b}^L\}$ may be expanded and written in analogy to equation \eqref{Waves-SD-LPbrkt}.  The Lie--Poisson bracket for the motion equations of the fluid component in \eqref{Fluid-SD-LPbrkt} is defined on the dual of the semidirect-product Lie algebra $\mathfrak{X} \circledS  (\Lambda^0\otimes \Lambda^3$) of vector fields, $X\in\mathfrak{X}(M)$, acting on the direct sum of functions $f\in\Lambda^0(M)$ and densities $D\in\Lambda^3(M)$ on the three=dimensional domain of flow, $M$, 
The dual coordinates are: the 1-form density, $\ob{m}=\ob{\bf m}\cdot d {\bf x}\otimes d^3x$, dual to vector fields; the advected density, $a_1=\Dt\,d^3x$, dual to functions;
and the advected scalar function, $a_2=\bb^L$, dual to densities. 

This means that the Lie-Poisson bracket in equation \eqref{Fluid-SD-LPbrkt}  may be written as
\begin{align}
\begin{split}
\Big\{F, H \Big\}(\ob{m}, a_1,a_2)&= 
\sum_{i=1}^2\SCP{( \ob{m}, a_i ) }{\left[ \left(\frac{\de F}{\de \ob{m}}\,,\, \frac{\de F}{\de a_i}\right)
\,,\,
\left(\frac{\de H}{\de \ob{m}}\,,\, \frac{\de H}{\de a_i}\right) \right] }_{\mathfrak{X},V}
\\ \\&=
\SCP{\ob{m}}{\left[\frac{\de F}{\de \ob{m}}\,,\, \frac{\de H}{\de \ob{m}} \right]}_\mathfrak{X}
+  \sum_{i=1}^2 \left(\SCP{\L_{\frac{\de F}{\de \ob{m}}} a_i}{\frac{\de H}{\de a_i}}_V
- \SCP{\L_{\frac{\de H}{\de \ob{m}}} a_i}{\frac{\de F}{\de a_i}}_V \right).
\end{split}
  \label{Fluid-SD-LPbrkt1}
\end{align}
The corresponding forms of the fluid equations in \eqref{Fluid-SD-LPbrkt} may then be written in terms of Lie derivatives are
\begin{align}
\begin{split}
\big(\p_t + \mathcal{L}_{\ob{u}^{L} }\big) \big(\ob{\bf m}\cdot d\bx \otimes d^3x \big)
&=  - \big( \Dt\,d^3x\big) d \Pi_{tot} + \big( \Dt\,d^3x\big)gz\,d\bb^L
\,,\\
\big(\p_t + \mathcal{L}_{\ob{u}^{L} }\big) \big( \Dt\,d^3x\big) &= 0
\,,\\
\big(\p_t + \mathcal{L}_{\ob{u}^{L} }\big)\bb^L &= 0
\,.\label{Fluid-LieDerivForm}
\end{split}
\end{align}
Thus, the particle momentum density, mass density and buoyancy are all transported by the same Lagrangian mean velocity.

The geometric similarities pervading the equations for the dynamics of the wave and material components  of the WCI system argues that it should be treated as a two-fluid system, e.g., as for $He$II.  
If so, then one should note that, just as for $He$II, the two fluids interpenetrate one another, since the wave and material properties are transported at different velocities. The material component of the GLM fluid is transported at the Lagrangian mean velocity, $\ob{\bf u}^L$, while the wave component of the GLM fluid is transported at the sum of velocities, $\ob{\bf u}^L+ \ob{\bm v}_G$.

The Lie--Poisson bracket for the WCI system is the sum of two Lie--Poisson brackets. That is, the Lie--Poisson bracket for WWCI is dual to the direct-sum Lie algebra 
\begin{align}
\mathfrak{G}=
\mathfrak{X} \circledS  \Lambda^0 \oplus \mathfrak{X} \circledS  (\Lambda^0\otimes \Lambda^3),
\label{Direct-sum}
\end{align}
whose dual coordinates have been identified in detail above. The direct-sum Lie algebra structure in \eqref{Direct-sum} means that the Lie--Poisson brackets among the wave quantities in \eqref{Waves-SD-LPbrkt} and material quantities in \eqref{Fluid-SD-LPbrkt} all vanish. 
However, as we saw in equation \eqref{GLM-Bouss-eqns2}, the fluid motion equation for the combined momentum density $\overline{\mathbf{m}} = \Dt(\ob{\mathbf{u}}^L + \ob{\mathbf{R}}^L(\mathbf{x})) - {\pb}$ will also be affected by the wave pseudomomentum ${\pb}$ equation, via a type of Lorentz force reminiscent of the `vortex force' in the Craik--Leibovich theory, except that the Stokes mean drift velocity $\ob{\bf u}^S$ in the CL theory will be replaced by the pseudovelocity $\ob{\bf v} = \ob{\bf p}/\Dt$ in equation \eqref{pvel-def}. The corresponding Lie--Poisson structure can be obtained by a linear change of variables.

\end{document}